\documentclass{IEEEtran}

\usepackage{algorithm,algorithmic}
\usepackage[utf8]{inputenc}
\usepackage[T1]{fontenc}
\usepackage{tabularx}
\usepackage{placeins}
\usepackage{amsmath}
\usepackage{graphicx}
\usepackage{comment}
\usepackage[breaklinks=true]{hyperref}
\usepackage{multirow}
\usepackage{breakurl}
\usepackage{lettrine}
\usepackage{mathtools}
\usepackage{listings}
\usepackage{makecell}
\usepackage[table]{xcolor}
\usepackage[breaklinks=true]{hyperref}
\usepackage{xurl}
\usepackage[center]{caption} 

\lstdefinelanguage{script}{
  keywords={},
  keywordstyle=\color{blue}\bfseries,
  ndkeywordstyle=\color{darkgray}\bfseries,
  identifierstyle=\color{black},
  commentstyle=\color{purple}\ttfamily,
  stringstyle=\color{red}\ttfamily,
  sensitive=true
}

\date{May 2020}

\begin{document}

\title{Review and Critical Analysis of Privacy-preserving Infection Tracking and Contact Tracing}
\author{\IEEEauthorblockN{William J Buchanan\IEEEauthorrefmark{1}, Muhammad Ali Imran\IEEEauthorrefmark{2}, Masood Ur-Rehman\IEEEauthorrefmark{2}, Lei Zhang\IEEEauthorrefmark{2}, Qammer H. Abbasi\IEEEauthorrefmark{2}, Christos Chrysoulas\IEEEauthorrefmark{1}, David Haynes\IEEEauthorrefmark{1}, Nikolaos Pitropakis\IEEEauthorrefmark{1}, Pavlos Papadopoulos\IEEEauthorrefmark{1}  
}\\

\IEEEauthorblockA{\IEEEauthorrefmark{1}Blockpass ID Lab, School of Computing, Edinburgh Napier University}\\
\IEEEauthorblockA{\IEEEauthorrefmark{2}Communication, Sensing and Imaging, James Watt School of Engineering, University of Glasgow}

}
\maketitle

\begin{abstract}
The outbreak of viruses have necessitated contact tracing and infection tracking methods. Despite various efforts, there is currently no standard scheme for the tracing and tracking. Many nations of the world have therefore, developed their own ways where carriers of disease could be tracked and their contacts traced. These are generalized methods developed either in a distributed manner giving citizens control of their identity or in a centralised manner where a health authority gathers data on those who are carriers. This paper outlines some of the most significant approaches that have been established for contact tracing around the world. A comprehensive review on the key enabling methods used to realise the infrastructure around these infection tracking and contact tracing methods is also presented and recommendations are made for the most effective way to develop such a practice.

\end{abstract}


\section{Introduction}

{In December 2019, the Wuhan area of China was the first in the world to experience the spread of novel Coronavirus Disease 2019 (COVID-19). Zhou et al. \cite{zhou2020clinical} found that in two hospitals in the area, of those who were discharged and had died, 48\% had a co-morbidity with hypertension (30\%) and coronary heart disease (8\%) being the most common.} It has since spread to many regions of the world. Horton \cite{horton2020offline} states that the Contain–Delay–Mitigate–Research strategy of the UK government failed as they initially did not test every suspected case, isolate, quarantine and trace their contacts. Scale of the crisis is highlighted with a delay of over three months on non-urgent surgeries \cite{iacobucci2020covid}, and where there are worries around the mental health of those dealing with COVID-19 patients \cite{greenberg2020managing}.

{Heymann et al. \cite{heymann2020covid} specifies that close monitoring is required in order to match public health approaches to their social acceptance and stresses that there is a need of strong communication methods for self protection, identification of symptoms and seeking the treatment. The COVID-19 pandemic has necessitated a contact tracing system which can be used to identify an infected individual, and then trace the people who have been in contact with that person. This provides ways to contain the spread of COVID-19 by identifying the infected cases and their contacts and enforcing appropriate restriction of self-isolation or quarantine.} It is likely that a contact tracing system will require the support of human contact tracers, thus making necessary the usage of a mobile phone application because the mobile phones are the only devices used frequently while supporting a wide range of functionalities. {The key aspects of a contract tracing application are listed below:}

\begin{itemize}
\item \textbf{Centralised or distributed?} A distributed infrastructure allows users to determine the other people that they have been in contact with, whereas a centralised approach uses a central server to store location tracing information. 
\item \textbf{Proximity based or GPS?} This either involves using Bluetooth methods to track whether a person has been in close proximity to another person, or where GPS location is stored for a user.
\item \textbf{Privacy-enhanced methods?} This involves the methods used to identify the user and their history of contacts.
\item \textbf{Open or closed source?} An important aspect with contact tracing is whether the methods are open source. If this is the case, then it is extremely important if they can be peer-reviewed and by whom. Within a closed source system, there is the uncertainty that for the bugs to be discovered, it usually takes longer periods of time while the products are already commercially used. 
\end{itemize}

Hellewell et al. \cite{hellewell2020feasibility} analysed how well contact tracing could be used to suppress the spread of COVID-19. For this, they used the reproduction number (R0), the delay from symptom to isolation, contract tracing probability, transmission before symptom offset, and the percentage of sub-clinical infections. 



Privacy plays a {vital} role within a contact tracing infrastructure. One quote with \cite{raskar2020apps} defines:
\begin{quote}
``Some of my patients were more afraid of being blamed than dying of the virus"\\
-Lee Su-young, Psychiatrist at Myongji Hospital, South Korea
\end{quote}

The rest of the paper {is} structured as follows: Section II describes the underlying contact-tracing technologies, while Section III analyzes all the proposed methods. Section IV discusses the DP-3T approach. Section V provides a view on the range of attacks on contact tracing methods {while Section VI provides a critical analysis on the feasibility of creating a privacy-preserving contact tracing application. Section VII concludes the work.} 

\section{Underlying technology}

The fast spread of COVID-19 throughout the world caused by severe acute respiratory syndrome has made it an unprecedented national health crisis \cite{alimadadi2020artificial}. It has brought health services under colossal pressure and necessitates for novel solutions to combat the spread. Contact tracing is one of the crucial interventions that public health professionals rely on in managing the early stages of disease outbreaks.

In the public health realm, the process of identifying an infected patient, listing and following the people who may have exposed themselves to the infection by coming into contact with them is termed as contact tracing. The process is reckoned to be an effective tool to prevent the spread of infection at a faster rate through timely provision of appropriate care to these people \cite{WHO}.  

Despite proving to be very useful in preventing the spread of a disease, the performance of conventional contact-tracing techniques (interviewing each patient and contacting people that have been exposed to the patient) is often inadequate in urban areas and during disease peaks \cite{Swanson2018contact}. Germany was able to hold off the disease for a few weeks by using manual contact tracing and moving COVID-19-positive patients to quarantine \cite{Spiegel}, but the effectiveness of contact tracing relies on the faster growth of identified cases than the number of new infections \cite{Eames2003contact} which is not possible in manual tracing. 

This limitation is stressed out by the exponential spread of the disease enabled by a higher population density and frequent movement of urban residents. Digital tools utilising existing technologies to gather information on the spread, key symptoms, and means the virus is employing to transfer are reckoned to be an effective response. Contact tracing is one of the key digital techniques that can not only enable the authorities to keep a track of the viral spread but can also play a pivotal role in identifying the potential carriers of the virus due to coming in contact with an identified patient. 29 countries around the world are now using mobile data to help with contact tracing COVID-19 cases \cite{QZ}.

This section summarises the communications standards and methods that would best support the contact tracing approach and options already progressing that may help us get further faster.

\subsection{Broadcast, selected broadcast, unicast and participatory methods} 
The methods involved with COVID-19 tracing typically split into: (a) crowd-sourced applications; (b) self-reporting systems; (c) centralised contact tracing; and (d) decentralised contact tracing. The importance of testing is underlined by Beeching, who outlines \cite{beeching2020covid}:

\begin{quote}
“Test, test, test” is the key to controlling the spread of SARS-CoV-2 and its clinical manifestation, covid-19, according to the World Health Organisation. However, three months after notification of the novel coronavirus infection in China, there is inadequate access to appropriate diagnostic tests globally and confusion among healthcare professionals and the public about prioritisation of testing and interpretation of results.
\end{quote}

\begin{figure}
\centering\includegraphics[width=0.45\textwidth]{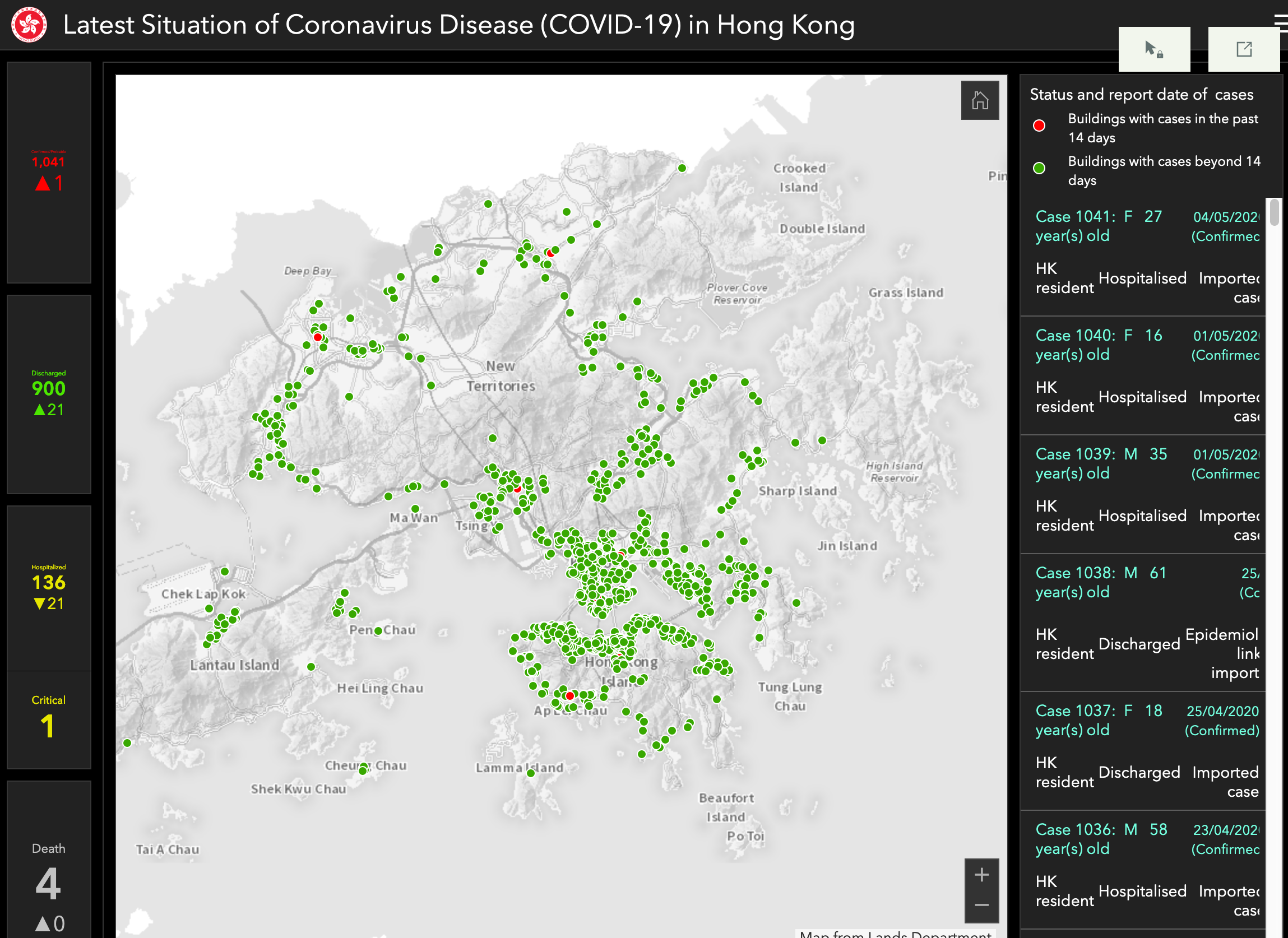}
\caption{Novel Coronavirus Infection in Hong Kong using ArcGIS}\label{fig_hong}
\end{figure}

Raskar et al. \cite{raskar2020apps} reviewed the risks around contact-tracing and defines the methods of broadcast, selected broadcast, unicast, and participatory methods. With \textbf{broadcasting}, the governments share the location of those who have been proven to be infected. Singapore and Hong Kong (Figure \ref{fig_hong}) have a detailed map of infected cases, while South Korea uses SMS messages about those who have tested positive.

With \textbf{selective broadcasting}, governments do not send out the information to the public about carriers and send only to a selected group, such as people within a geographical area. In this case, users will register for location information, through a specific App, or for their phone numbers. In this case, it may only send you information related to a carrier being in a certain location. 

For \textbf{unicast}, people are informed when a carrier moves into contact with another person. This method was used in China for those who were suspected to be at risk. The message is then targeted to the person that goes close to a suspected carrier. Generally, it is poor in terms of privacy, and it requires detailed surveillance of citizens.

In terms of citizen engagement, the methods of \textbf{participatory sharing} is one of the strongest, as users share their locations with a central authority. They may also share information that allows others to understand the risks that they, and others, face. Unfortunately, this method may be open to abuse from fraudulent entities.

Raskar et al. \cite{raskar2020apps} have also analysed the risks of these methods in based on (Table \ref{fig_diff}):

\begin{itemize}
\item Accuracy: In this, unicast method has the lowest risk.
\item Adoption: In this, broadcast method has the lower risk. 
\item Privacy: In this, broadcast, selected broadcast and participatory methods have the lower risk.
\item Consent: In this, practices vary greatly. For participatory, full consent is required. 
\item Systemic challenges: In this category we are dealing with issues like: fraud and abuse, and the security of information. The risks are considered high in all the categories. 
\end{itemize}

\renewcommand\theadfont{\bfseries}
\begin{table*}
    \caption{Risks and challenges of contact-tracing technological approaches (reproduced from \cite{raskar2020apps})}
    \label{fig_diff}
    \centering
     \setlength{\leftmargini}{0.4cm}
    \begin{tabular}{| m{2.5cm} | m{2.5cm} | m{2.5cm} | m{2.5cm} | m{2.5cm} | m{2.5cm} |}
        \hline
        \thead{Intervention} & \thead{Broadcast} & \thead{Selected Broadcast} & \thead{Unicast} & \thead{Participatory} & \thead{PrivateKit}\\
        \hline
        
        \textbf{Accuracy}  & Limited & Limited     & High    & Limited    & High \\\hline
        
        \textbf{Adoption}    & High      & Medium    & Medium     & Low   & Medium              
        \\\hline

        \textbf{Privacy risks for carriers} & Significant  & Moderate     & Moderate  & Significant  & Moderate to Low                           \\\hline
        
        \textbf{Private risks for local businesses} & Significant   & Significant  & Moderate     & Significant      & Moderate to Low   
        \\\hline

        \textbf{Privacy risks for users }   & Privacy-protected   & Privacy at risk      & No privacy     & Privacy protected    & Privacy protected     \\\hline

        \textbf{Privacy risks for non-users}   & Privacy at risk if carriers are identified & Privacy at risk if carriers are identified & Privacy at risk if carriers are identified & Privacy at risk if carriers are identified   & Privacy at risk if carriers are identified
        \\ \hline

        \textbf{Consent of carries}   & Practices vary & Practices vary & Practices vary, often with little or no consent & Full consent  & Full consent
        \\ \hline

        \textbf{Misinformation and panic}  & High risk & Medium risk & Medium risk   & High risk & Medium risk
        \\ \hline

        \textbf{Security of information}  & Low-to-medium risk & Low-to-medium risk & High risk    & Low risk & Low-to-medium risk
        \\ \hline

    \end{tabular}
\end{table*}

\begin{figure*}
\centering\includegraphics[width=0.8\textwidth]{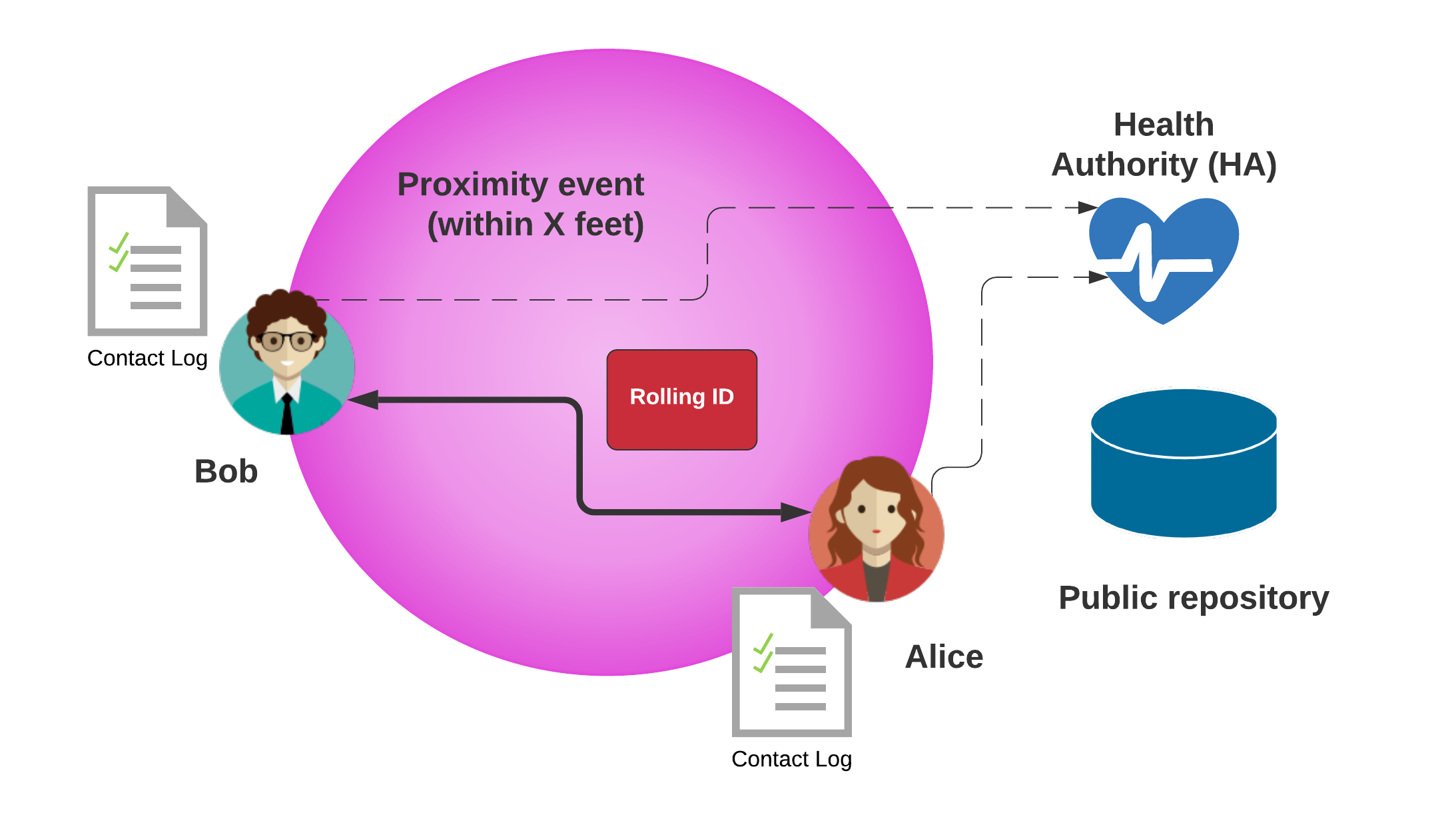}
\caption{Basic model of contact tracing}\label{fig_bobaliceha}
\end{figure*}

Contact tracing is seen to be a part of reducing the spread of the COVID-19 disease, and many countries of the world have moved to integrate technical solutions for contact tracing. These methods are mainly Bluetooth-based {where a Bluetooth beacon} is sent between Bob and Alice when they are within a given proximity and for a minimum amount of exposure time. At the core of these methods is whether the {approach} is \textbf{centralized} (where those infected are matched on a central server), or \textbf{decentralized} (where {individuals} can do their own matching with full consent). Basically, there are three main entities involved: Bob (who is infected), Alice (who is in contact with Bob) and the HA (Health Authority), as shown in Figure \ref{fig_bobaliceha}. We may also introduce Grace, the government official, and Eve, the eavesdropper.

Within a centralized system, Bob and Alice are assigned identifiers that the HA can match whenever Bob and Alice are in contact. {This is matched through a privacy-preserving rolling ID which only the HA can match back to Bob and Alice. Once matched, the HA can then inform Bob and Alice that they have been in contract.} In a decentralized system, Bob and Alice send the rolling IDs they receive, and Bob can identify that he has COVID-19. The HA can then keep an ID resolver so that Alice can determine when she has been in contact with Bob. In this way, the HA does not know about the contact between Bob and Alice, but Alice will. 

\section{Contact Tracing Apps for COVID-19}

Almost three quarters (72.6 \%) of internet users will access the web solely via their smartphones by 2025, which is equivalent to nearly 3.7 billion people \cite{CNBC}. The proliferation of mobile devices presents a new opportunity for overcoming the challenges faced by conventional contact tracing techniques in terms of identifying, monitoring and informing about the spread of a pathogen in densely populated areas, as is the case with COVID-19. A famous example is China, which relied on an elaborate surveillance architecture to actively monitor the location of its citizens using live data and mobility history to enforce self-isolation and conduct contact-tracing. Several other countries, like Korea, Singapore, Israel, Iran, and Russia \cite{BBC_China}, have built solutions around the Chinese model. However, these attempts have led to criticisms on privacy and data protection. 

A variety of different groups around the globe are working on the same lines to develop a contact tracing app. {The following subsections outline some of the key technologies either defined by organisations, country or geographical region.}

\subsection{Apple and Google}
Apple and Google have worked together to create an API integration for Bluetooth to track physical proximity between phones. If someone later receives a positive COVID-19 diagnosis, they can report it through the App, and any users who have been in recent contact with the infected person will receive a notification. The system is Bluetooth-only, fully opt-in, collects no location data from users, and no data at all from anyone without a positive COVID-19 diagnosis. Apple and Google chose perhaps the most privacy-friendly of the many different schemes that could allow automated smartphone contact tracing \cite{Wired_Amazon_Apple}. 

\begin{figure*}
\centering\includegraphics[width=0.8\textwidth]{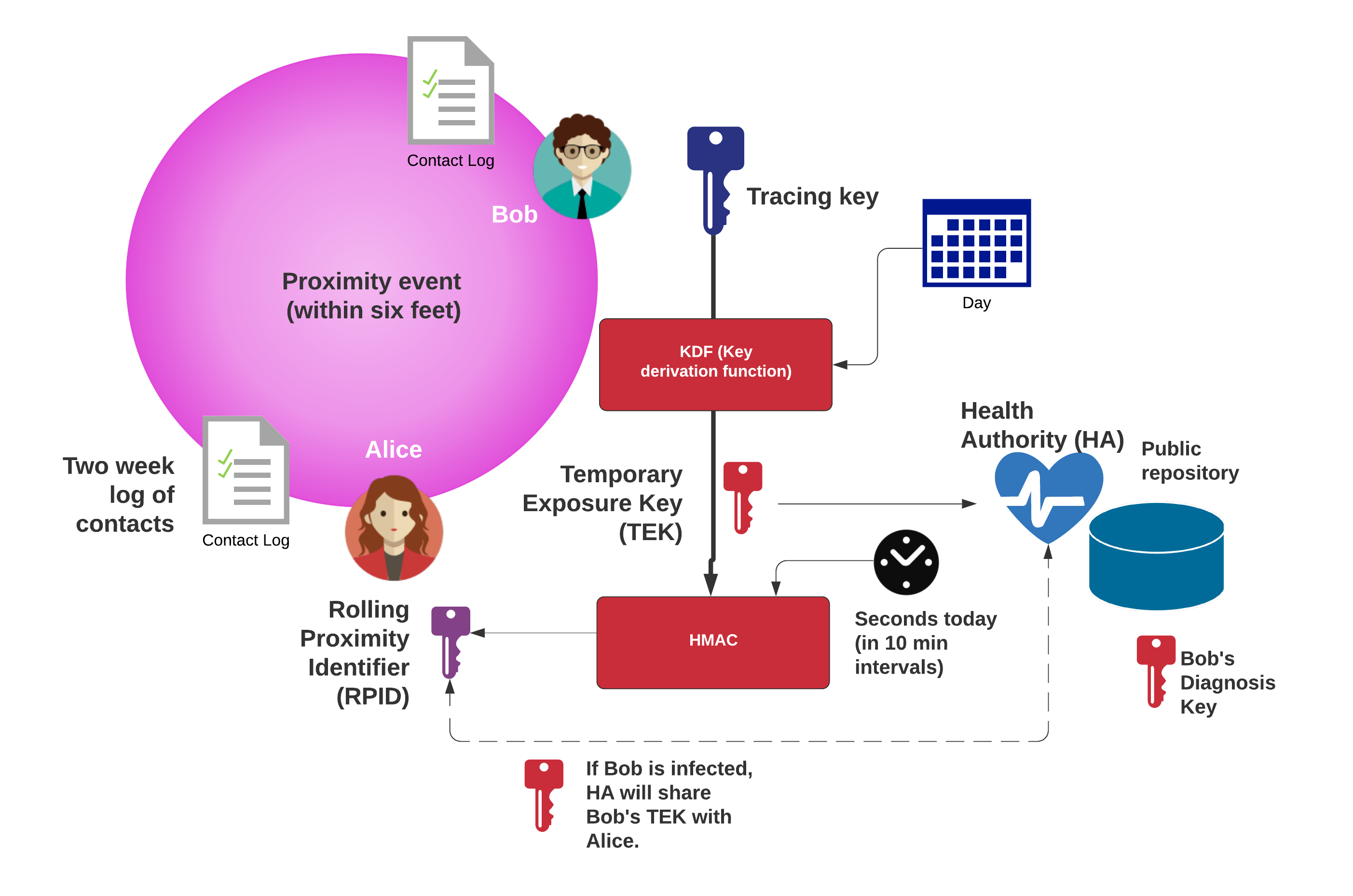}
\caption{Google/Apple ID tracing}\label{fig_apple01}
\end{figure*}

The Apple/Google method works by Bob generating a unique 256-bit tracing key for his phone — and where this key must be kept secret (Figure \ref{fig_apple01}). Every day he then creates a daily tracing key (diagnosis key), by creating a hash from the tracing key and the current day. From this hash, it should not be possible to determine his tracing key (as it is generated from the random 256-bit tracing key). {Every 10 minutes, he  creates a rolling ID key which is an HMAC identifier (a signed hash) of his daily tracing key and a counter for the number of 10 minutes that have passed that day.}

When Alice comes into contact with Bob, she will receive her rolling ID through a Bluetooth Advertisement, and could then pass that back to the Health Authority (HA). The HA cannot correlate Bob from the rolling ID, and whether he has COVID-19 or not. Alice is just blindly sending it back to Trent, in order for the HA to track the contact or not. In order to preserve privacy, the HA should only track if Alice has been proven to be COVID-19 positive.

{Once Bob has been proven to have COVID-19, he will send the daily tracing key to the HA, who can then match all the rolling ID keys to his identity. This will only happen for one day. As he must send these keys every day, the key feature on the phone will make the decision to send the key daily or not. To enhance security, this design has been updated to integrate AES encryption.} 

\subsection{The United Kingdom}    
In the UK, KCL, Guys and St Thomas’ Hospitals in partnership with ZOE Global Ltd have proposed C-19 COVID Symptom Tracker \cite{Zoe}. The data is collected through daily self-reporting of a volunteer user and analysed by machine learning and data science methodologies to predict high-risk areas in the UK while understanding symptoms, and the propagation of the virus. However, the specific application is not widely being adopted because it does not offer any real-time tracing.

\begin{figure}
\centering\includegraphics[width=0.5\textwidth]{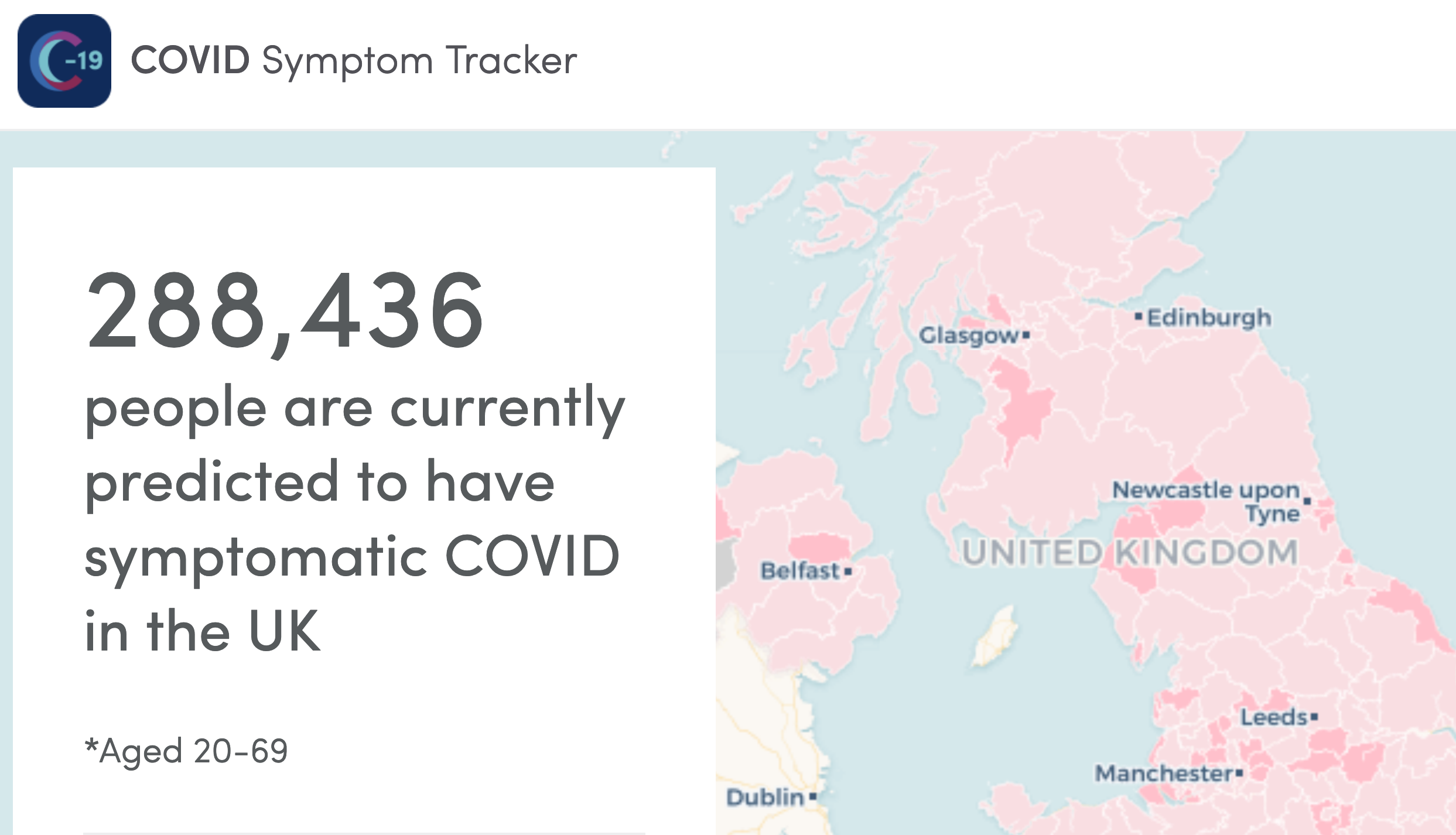}
\caption{Covid.joinzoe.com}\label{fig_job}
\end{figure}

Susan Major \cite{mayor2020covid} defines that around 700,000 people installed the COVID Symptom Tracker app (covid.joinzoe.com) within the first day of its release (Figure \ref{fig_job}). The research team are critical that the current focus is on deaths rather than tracking the spread of the disease in the percentage of the population who were symptomatic (estimated between 5\% and 60\%). The App includes information related to age, sex, height, weight, and postcode and lists any chronic health conditions. Along with this users are asked about their current symptoms including whether they have a fever, persistent cough, unusual fatigue, shortness of breath, diarrhoea, confusion, disorientation or drowsiness, and loss of appetite. Those reporting symptoms are then sent a home testing kit, and where the data received is then used to report whether they have COVID-19 or not. All of the data collected is anonymised and will be provided back - free of charge - to researchers. 

Babylon COVID-19 Care Assistant is developed by Babylon Health (a private healthcare provider) \cite{Babylon}. It is a separate section of an existing Artificial Intelligence (AI) based App. As an existing user logs in, they are asked at the beginning of the triage if they are concerned about coronavirus. If they answer yes, they are diverted to the COVID-19 triage, which follows the same criteria of NHS111. The NHS representative provides people with updated information about coronavirus, allows them to log their symptoms, helps them get appropriate assistance and advice to help them with not spreading the virus wider. The App includes a live chat run by clinical support staff and overseen by doctors.

\begin{figure*}
\centering\includegraphics[width=0.7\textwidth]{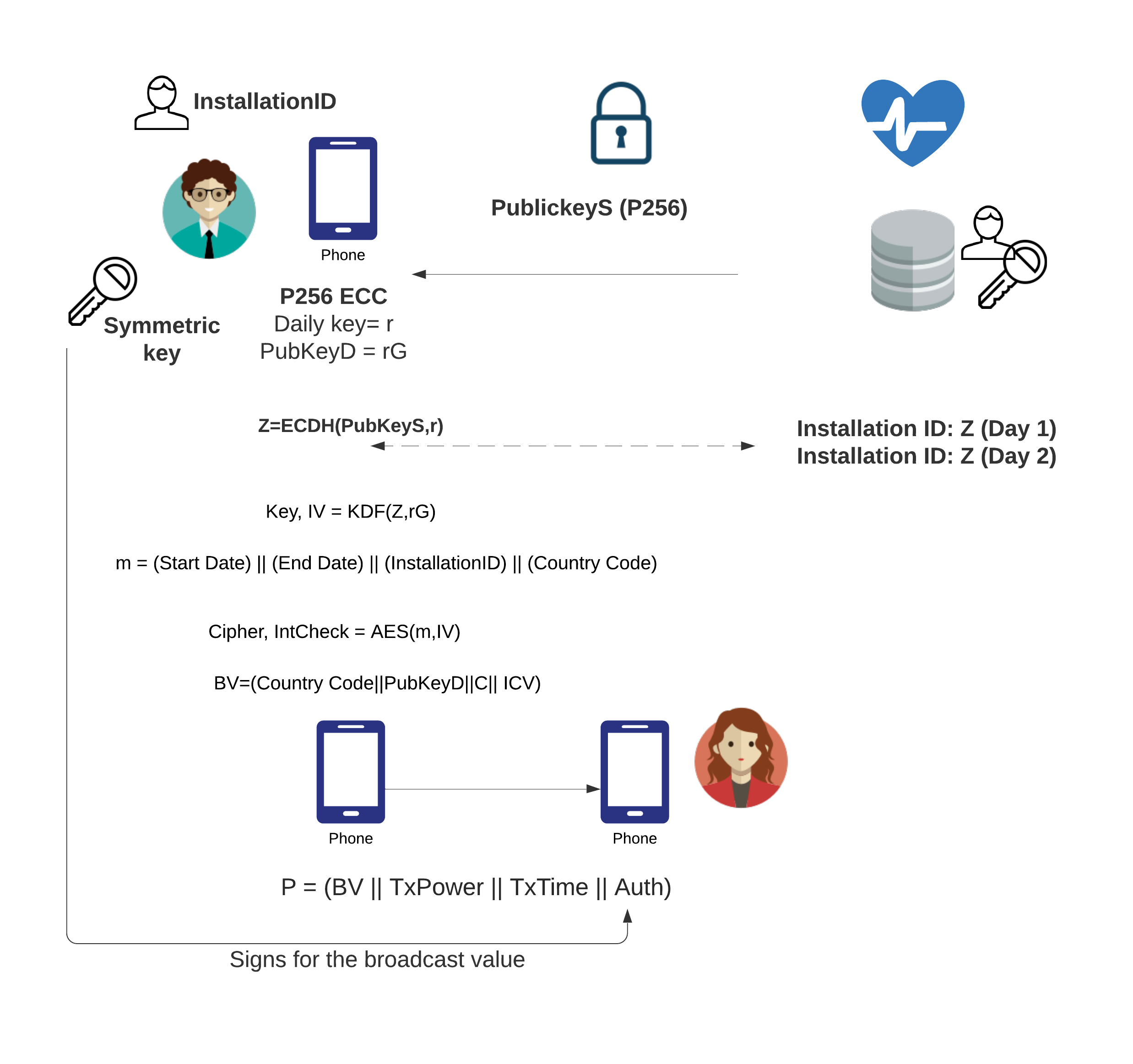}
\caption{ECIES Encryption with the NHSX App}\label{fig_nhsx01}
\end{figure*}

{The UK government developed an NHSX App based on Low Energy Bluetooth}. Once installed, the App allowed the logging of the encrypted information of the users operating in the close proximity of the host into a database \cite{NHSX}. A COVID-19 positive case would lead to the App alerting everyone who was noted to be in the vicinity of the detected user. Within the NHSX App, we use elliptic curve cryptography, and use a off-line key exchange method known as ECIES (Elliptic Curve Integrated Encryption Scheme). {As illustrated in Figure \ref{fig_nhsx01}, Bob receives an InstallationID, the public key of the HA ($PubKeyS$) and a symmetric key from the HA.} Every day he then creates a daily public key pair:

\begin{equation}
DailyPriv = r
\end{equation}
\begin{equation}
DailyPub = rG
\end{equation}

\begin{figure*}
\centering\includegraphics[width=0.7\textwidth]{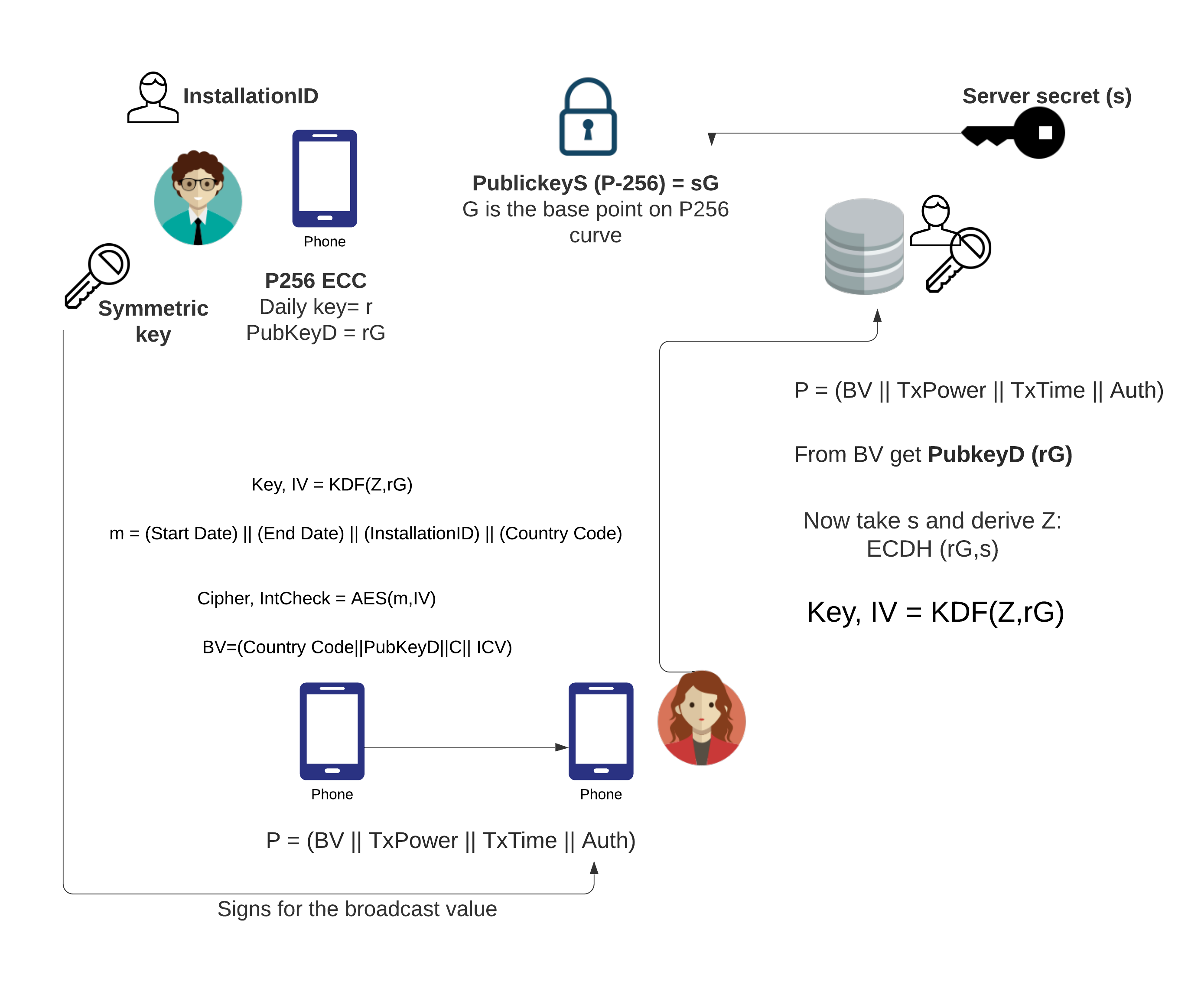}
\caption{Decrypting Bob's ID with the NHSX App}\label{fig_nhsx02}
\end{figure*}

{Where} $r$ is a random value and $G$ is the base point on the chosen elliptic curve. Using the public key of the HA and his own daily private key ($r$), Bob generates a secret value ($Z$). This value can also be regenerated at the HA with the private key of the server ($PrivS$) and Bob's daily public key ($rG$). From $Z$, Bob generates the encryption key which will protect the InstallationID for Bob. The symmetric key passed is used to sign for the Bluetooth beacon. Finally, Bob adds his daily public key to the beacon. This public key will be used to regenerate the secret value ($Z$) at the HA, and thus generate the same encryption key. Figure \ref{fig_nhsx02} outlines the decryption process using Bob's daily public key and the HA's private key.


The approach is thus centralised, and where Bob's identity is created by the HA, and then matched back. One possible weakness is where the private key of the HA is leaked, and which will allow all of the derived keys to be generated. Along with this, Bob's public key will be exposed for a day (rather than 10 minutes in the case of the Apple/Google contact tracing method). It could thus be possible to trace Bob for one day by monitoring his Bluetooth beacons and matching his daily public key. After trails on the Isle of Wright, the App was found to be good at determining distances between devices, but it only found 4\%of Apple devices and 75\% of Android devices \cite{NHSXfailure}. This led to cancellation of the centralised approach, and a move towards the hybrid approach.

{BeepTrace \cite{xu2020beeptrace} is a blockchain-enabled fully distributed privacy-preserving solution for COVID-19 contact tracing. in this approach, blockchain is adopted to bridge the user/patient and the authorised solvers to desensitize the user ID and location information. Compared with the dominating active mode that adopts Bluetooth or similar technologies to exchange information among contacted users, BeepTrace proposes a passive model by using GPS or similar without direct information exchange at the end user side.} 

\subsection{The European Union}
The European Pan-European Privacy Preserving Proximity Tracing Initiative (PEPP-PT) proposes an open source Bluetooth based platform sharing software, standards and services that can be utilised for the development of COVID-19 contact tracing Apps. Each national health authority can tweak the software according to its own policies and processes. The software aims at measuring proximity data and alerting the traced contacts of a user if detected positive to COVID-19 while adhering to privacy. As an EU initiative, it has a wide approach across national borders \cite{PePP}.

One of the methods that has been published which tries to address this balance is Pan-European Privacy Preserving Proximity Tracing (PEPP-PT).  With this, a device requests and ID from the tracking service (Trent) and is given a one-time anonymised ID (and which includes an obfuscation of the country ID). {It then use Bluetooth beacons and possibly Wi-Fi to discover and identify neighbours.} The method defined in PEPP-PT then uses the signal strength method to estimate the distance someone is away. {Note that this is not a GPS tracking method and will just give a circular radius around a person, and possibly amount of time that they were near another phone.}

The users must install the tracking application on their phone — possibly they must be forced to do this by their government — and then it will be used to track contacts between one phone and another.

The results are then sent back to Trent with a device identifier for Alice’s phone (possibly the Bluetooth MAC address). If the device is a registered device with an anonymous ID, it will send back its neighhour’s ID and an estimation of location, and also store this as in the history log.

Overall, there is no personal information stored, and the device just stores anonymised IDs. The history is then deleted when there is a test that the user of the device does not have SARS-Cov-2, but remains encrypted until there is a test to prove that they do not have the virus.

If the user has been proven to have it, the health authority registers the device with a TAN code, and with consent, they register onto a tracking system, and where they allow others they have been in contact with to be alerted to the possible threat of infection. There is no personal information stored. If the phones are from different countries (identified in the anonymous ID), there is an alert send to the health care provider in the other country.

\subsection{DP-3T}
The DP-3T is decentralised and open sourced \cite{troncoso2020decentralized}. It involves a collaboration of eight different countries. It general defines a number of objectives:

\begin{itemize}
\item Anonymous identifier donation. This uses a short-term  anonymous identifier (ID) and includes a measure of the Bluetooth signal strength.
\item Logging the proximity history. This only stores proximity information when within an epidemiologically sufficient proximity. No geo-location information is stored.
\item Usage of the proximity history. The data is stored in an encrypted form if the user has not been tested, or is clear. When tested positive, the Health Authority contacts the user, and provides a TAN code, and where the user then consents to reveal their history. 
\item Country-dependent trust service operation. If the phones are from different countries (identified in the anonymous ID), there is an alert send to the health care provider in the other country.

\end{itemize}
\subsection{France}
In France, Covidom App is developed to monitor COVID-19 patients who have been through Paris hospitals and either identified as COVID-19-positive or suspected of being infected) but do not require hospitalization but are staying at home \cite{RFI}. The application is based on a daily digital online questionnaire asking the patient about their respiration, heart rate and temperature. Depending on the response of the patient, the healthcare team is alerted and contacts the patient to adapt the follow-up and treatment. The app is voluntary and does not have any real time contact tracing capability.

\subsection{Poland}
In Poland, Home Quarantine App is developed and endorsed by the Government. Its use is mandatory for 14 days to people returning to Poland from abroad and for those who are COVID-19-positive \cite{BusinessInsider}. The app is based on Instagram and require the people to upload their selfies within 20 minutes of receiving an alert. Instagram’s geolocation and facial recognition capabilities are used to ensure that the people are adhering with self-quarantine. 

\subsection{Germany}
Germany has Corona Data Donation App that gathers the vital signs (pulse, temperature, sleep) of volunteers and analyse the probability of testing positive for COVID-19 using wearable technology \cite{Reuters}. An online interactive map is generated based on this information to depict the geographical spread of the virus.

\subsection{Russia}
The Russian Social Monitoring App tracks the self-isolated COVID-19 patients’ whereabouts through user’s calls, location, camera, storage, network information and other data \cite{NPR}.  
\subsection{China}
In China, Health Code App - developed by the Government, WeChat and Alipay - tracks people's symptoms and issues real-time individual health status, using a three-colour scheme (Green, Yellow, and Red) \cite{NYTimes}. Real-time location and tracing based on GPS, existing WeChat/Alipay payment system, mobile network and traffic data with advanced machine learning/big data analytic is employed to enable accurate detection and fast alerting. The app usage is compulsory and may lead to potential data/privacy issues as well as discriminating behaviours based on colour schemes.

\subsection{South Korea}
In South Korea, Corona 100m (Co100) App allows those who have been ordered not to leave home to stay in contact with case workers and report on their progress \cite{SmartCitiesWorld}. The App uses GPS to keep track of infected people’s location to make sure they are not breaking their quarantine. It also alerts users when they come within 100 metres of a location visited by an infected person. Machine learning/data science tools are however not used to track/trace travel history and make real-time alerts to the public. The app is not mandatory, and a user can opt out.

South Korea’s Ministry of Interior has also introduced a mandatory GPS based Self-quarantine Safety Protection App to support officials to monitor citizens in quarantine \cite{GeoSpatialWorld}. An alert is sent to both patient and case worker, if the patient leaves their quarantine zone. Citizens can self-report their symptoms. 

\subsection{Singapore}
Singapore have introduced TraceTogether, a Bluetooth based App that traces and identifies those who have been exposed to a COVID-19 infected person \cite{Tracetogether}. The scan history is stored locally. The participation is voluntary, and no real-time alerting and tracing is available. TraceTogether uses Bluetooth and keeps a track of all the contacts made within a 21-day period \cite{mccall2020shut}. It only stores contacts and not the actual locations of the phone. If the Ministry of Health requires the contact history, they ask the user for consent to share it. The logs are encrypted on the device, and only decrypted once the logs are uploaded to the Department of Health. Contract tracers then use the logs to match to those who the user has been in contact with. Bluetooth was selected due to the inability of GPS to locate accurately within buildings (as the GPS methods on phones only estimate within buildings). 

\subsection{India}
In India, AarogyaSetu App uses GPS and Bluetooth to track the people who have symptoms and identify people who have been in close proximity to them \cite{EconomicTimes}. The participation is voluntary, and the location and contact history stay on the device unless a user is COVID-19 infected in which case the person's data is sent to the cloud. The app traces travel/contact history but no real-time alerting is available. 

COVID-19 Quarantine Monitor Tamil Nadu App is also an Indian initiative that tracks a quarantined user. The voluntary used app enables live location tracking via GPS and generates alerts \cite{DeccanHerald}. The Apps in India have no underlying legal framework for privacy protections in place. 

\subsection{The USA}
\begin{figure}
\centering\includegraphics[width=0.45\textwidth]{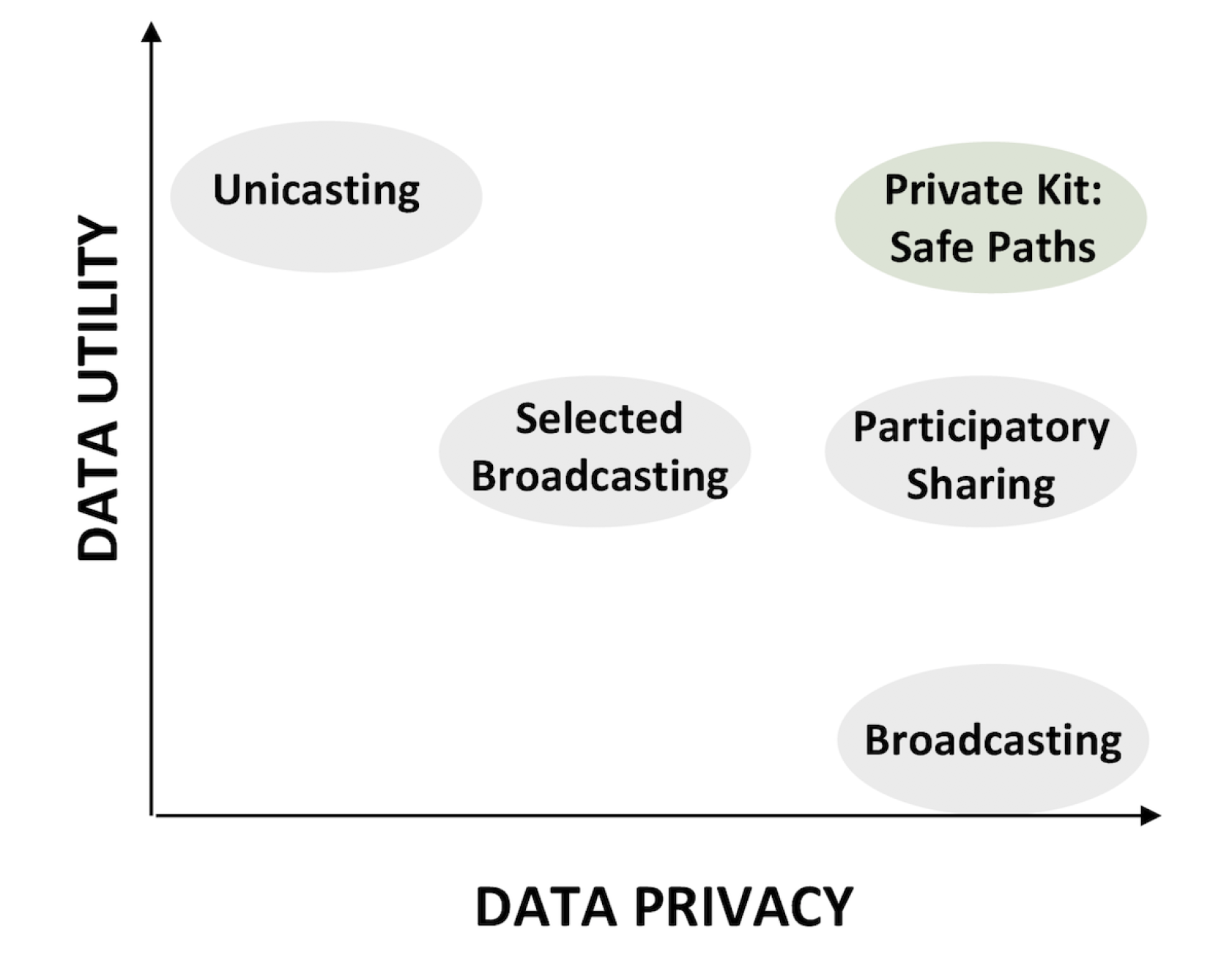}
\caption{Data Privacy and Data Utility}\label{fig_safe}
\end{figure}

In the USA, MIT developed Safe Paths App, which uses Bluetooth to track users and share locations between them \cite{SafePaths}. Safe Paths collects users' location data, keeping a time-stamped log every five minutes and is encrypted and stored locally. In total, 28 days of data can be stored in the app in under 100 kilobytes of space (that's less storage space than a single photo takes up). If a user is tested positive for COVID-19, they can share this data to health official by using a QR code, thereby facilitating contact-tracing \cite{PopularMechanics}. It also compares recent locations against the path of an infected person and alerts them of potential contact. SafePaths uses Bluetooth tracing and GPS methods \cite{cho2020contact}. Figure \ref{fig_safe} outlines that SafePaths has strong methdods of data privacy and data utility \cite{raskar2020apps}.

With SafePaths, the server limits clients to $N$ location points per exchange, and also limits the number of queries that a client can request every day. This stops an adversary mapping out a whole geographical area with continual requests. 

How We Feel App developed by Pinterest with the help from Harvard, Stanford, MIT, University of Maryland, Weill Cornell and the Howard Hughes Medical Institute, gathers the data on user’s health, age and zip code (information like name, phone number or email is not collected). The data is then aggregated and shared with researchers, public health professionals and doctors \cite{HowWeFeel}.

MIT is developing a Bluetooth based Private Automatic Contact Tracing App, where individuals enable their phone to continuously send out random data strings and keep a log of those from other participating devices it has encountered \cite{MobiHealthNews}. When a user is diagnosed with COVID-19, they would receive a QR code notifying a cloud system of their status. All other participants in the system would be able to scan the collective logs and would be warned of a potential (but still anonymous) COVID-19 contact.

\begin{figure}
\centering\includegraphics[width=0.5\textwidth]{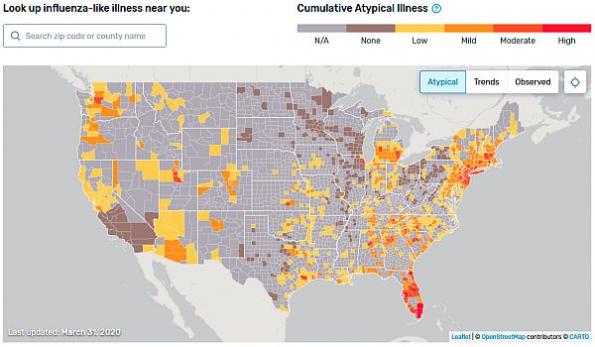}
\caption{Fever map in the US}\label{fig_kin}
\end{figure}

The US Health Weather Map App is created by Kinsa Insights and Oregon State University. It is currently being used to track \emph{typical illness levels} such as self-reported fevers (Figure \ref{fig_kin}). This type of application could be used to crowd source population health information, and on a scale that public health authorities would struggle with.

\subsection{Iran}
The Iranian Ministry of Health developed a contact tracing App requiring the user to register using his mobile number. The App uses GPS data to identify the user’s location. It also requests permission for identifying the user’s activity and shares the self-declared attributes of the user with the server including gender, name, height, and weight \cite{Avast}. Though removed from the Google Play Store, the App is still available from other application stores \cite{ZDNet}.

\subsection{Israel}
The ‘Hamagen’ App from Israel’s Health Ministry collects location history of the user using GPS in the background and compares the user’s movement with the health ministry’s data. If a user has come in close contact with a COVID-19-positive user, an alert is sent to the user directing them to a website containing details on further actions. Though it is stressed that all the information is stored on the user’s smartphone, the App comes alongside controversial temporary powers granted to Shin Bet security agency allowing them to track the movements of smartphones users via their devices and sending alerts to those who may have been exposed to COVID-19 being in contact with a confirmed infected user \cite{TimesofIsrael}.

\subsection{Australia}
The Australian government has launched a Bluetooth based contact tracing App COVIDSafe. The voluntary participation registers a user with name, age range, postcode and phone number. The system creates a unique encrypted reference code for the user. When the App recognises another device having the COVIDSafe App, it notes the date, time, distance and duration of the contact and the other user’s reference code in the form of encrypted data in the user’s phone \cite{Health_Australia}. When a user is detected as COVID-19-positive, this App data is acquired by the health officials who the send the alerts of possible exposure to the people traced as being in contact with that user.  

\subsection{EPIC}
\begin{figure*}
\centering\includegraphics[width=0.8\textwidth]{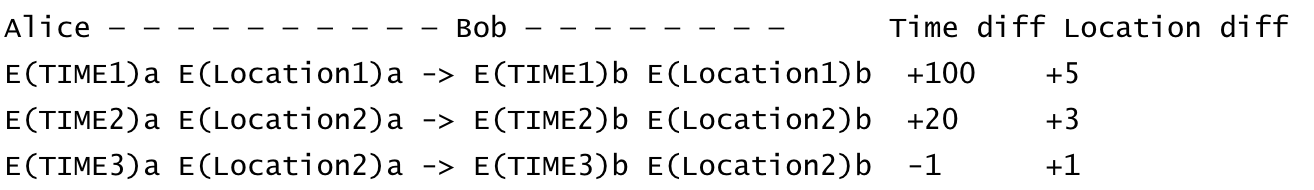}
\caption{Google/Apple ID tracing}\label{fig_id}
\end{figure*}

The EPIC contact tracing system uses homomorphic encryption in matching users up for possible contacts in a defined time window \cite{altuwaiyan2018epic}. First, Alice defines data in time stamps and stores homomorphically encrypted timestamps for her location:

\begin{verbatim}
E(TIME1)a E(Location1)a
E(TIME2)a E(Location2)a
E(TIME3)a E(Location2)a
\end{verbatim}

{Where} E(TIMEx)a is the homomorphically encrypted timestamp value, and E(Locationx)a is the homomorphically encrypted location information. Then, Alice and Bob upload their homomorphically encrypted time stamp and location information to the HA (Health Authority), who stores these values:

\begin{verbatim}
E(TIME1)a E(Location1)a
E(TIME2)a E(Location2)a
E(TIME3)a E(Location2)a
E(TIME1)b E(Location1)b
E(TIME2)b E(Location2)b
E(TIME3)b E(Location2)b
\end{verbatim}

The HA cannot tell either the time stamp or the location information. Alice is now identified as having COVID-19, and the server can identify her encrypted values and runs a homomorphic difference on the timestamps and location (Figure \ref{fig_id}).

\begin{figure}
\centering\includegraphics[width=0.4\textwidth]{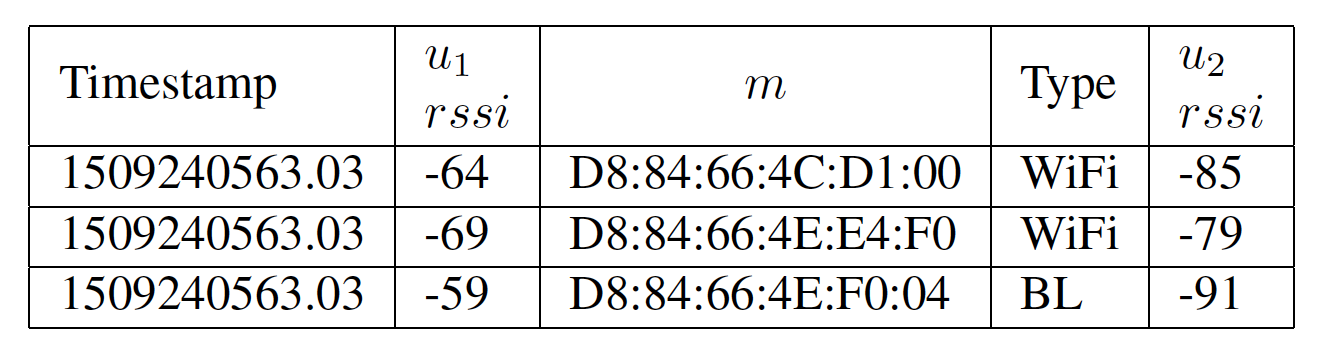}
\caption{Signal strength}\label{fig_sig}
\end{figure}

Here the HA cannot tell where Bob and Alice were and at what time, but they can tell that there was a match for a one-second difference and if they were one metre away from each other. In this way, Bob could be informed of a possible infection. Other information too is stored {which can} be used for the matching process, such as the device type, the SSID of the wireless access point that they connected to, and the RSSI (Received Signal Strength Indication), as shown in Figure \ref{fig_sig}.

\subsection{TraceSecure}
In TraceSecure \cite{bell2020tracesecure}, the authors proposed two private contact tracing methods using Bluetooth signals. The first one extends the TraceTogether application \cite{cho2020contact} by integrating a secure message-based protocol, while the second one incorporates a public key infrastructure that elaborates additive homomorphic encryption.

\subsection{Cryptographic Preservation of Privacy}
{Another approach based on GPS location data that incorporates strong encryption techniques, such as Private Set Intersection (PSI), has also been developed \cite{berke2020assessing}. The goal was the development of an infrastructure that not only promotes stronger privacy guarantees than the methods being adopted from the governments, but also feasible practically.} However, the performance overhead of a technique like this is being questioned \cite{bell2020tracesecure}.{Though the Epione \cite{trieu2020epione} solution uses lightweight cryptography to provide strong privacy, a real-world implementation has not been developed yet.}

\section{Key Enabling Technologies}
Location data is pivotal to any contact tracing solution. These solutions are based on the assumption that if two persons have shared a close proximity, they have contacted with each other. Artificial Intelligence based technologies such as facial recognition can be employed to reduce the number of false positives however; their limited availability restricts their usage. On the other hand, universal usage of mobile devices, smartphones and internet, GPS, Bluetooth beacons, {Wi-Fi, telecom cell towers} and social media can be effectively used to collect the user’s location data. GPS, Wi-Fi routers and cell towers provide absolute location data in the form of geolocation coordinates while Bluetooth pairing gives relative location data in the form of some reference description of the location, for example, both persons shared the same bus \cite{tang2020privacy}. 

Bluetooth tracing has emerged as the most suitable method for contact tracing in the backdrop of COVID-19 \cite{brack2020decentralized}. However, it has its own deficiencies that limits its capabilities. Methods such as private messaging for notifications of possible contacts after collecting Bluetooth IDs \cite{cho2020contact}, use of geolocation information \cite{NYTimes_Israel}, using Wi-Fi access, cellular network usage, social media, radio frequency identification and wearable devices have the potential to be used as contact tracing enablers. It is also possible to use smartphone built-in sensors such as gyroscope and magnetometer to correlate similar locations without revealing the actual coordinates where they occurred \cite{jeong2019smartphone}.

Contact tracing solutions does not have a uniform system architecture. Whilst countries rush to deploy the contact tracing apps, they raise a multitude of privacy and data protection issues. In this section, we discuss the technologies that can be effectively used to carry out contact tracing in the backdrop of COVID-19.

\subsection{Real Time Location Systems (RTLS)}
RTLS refers to any system that accurately determines an item or person’s location. RTLS is not a specific type of system or technology, but rather is a goal that can be accomplished with a variety of systems for locating and managing assets. An important aspect of RTLS is the time at which users are tracked, and this data can be used in different ways depending on the application. For example, some applications only need timestamps when a user passes through an area, while other RTLS applications require much more granular visibility, and entail that time data be updated constantly. An ideal RTLS can accurately locate, track, and manage assets, inventory, or people and help authorities to make knowledgeable decisions based on collected location data. RTLS is used across many industries including manufacturing, mining and healthcare industry. 

All RTLS applications will consist of a few basic components including a transponder, a receiver, and software to interpret the data. The complexity of the system, chosen technology, and scope of the application will determine the amount of hardware and software required to create the ideal RTLS.

\begin{itemize}
    \item \textbf{Global Navigation Satellite Systems (GNSS)}: Ever-growing demand of navigation and positioning facilities to be available in portable devices has made the GNSS an essential part of the modern communication applications, especially the mobile phones. A Federal Communications Commission (FCC) adoption to enhance the provision of emergency services by tracking a user’s location through his mobile also necessitates the integration of the GPS to the cellular phones \cite{miller1998adding}. The usage of GNSS for navigation-enabled smartphones is predicted to rise to 6.5 billion in 2020 \cite{GNSS_Market}.\\
    
    GNSS facilitates innovative tracking solutions, including the deployment of local geofences that trigger an alarm when a user leaves the perimeter. Global Positioning System (GPS) delivers the navigation and positioning services world-wide being the only fully functional satellite navigation system at present. The navigation systems are based on a fundamental positioning procedure where knowing the distance from an unknown location to a certain number of known locations, allows finding the coordinates of the unknown position. In the GPS, a number of satellites orbiting the earth provide the known locations while the position of the user on earth with a receiver is the unknown location \cite{logsdon2012navstar}.\\
    
    To determine 3-D position of the receiver, the principle of triangulation is used through the measurements of time delay between transmission and reception of each GPS radio signal transmitted by the GPS satellites. The distance between the user and the satellite is calculated from this time delay as the speed of signal (equals to the speed of light) is already known. The GPS signals also carry information about the location of the satellites. By determining the position of, and distance to at least three satellites, the GPS receiver can compute its position in terms of latitude, longitude and height (Figure \ref{fig_gps_1}). However, a fourth satellite is also required for a timing offset that occurs between the clock in the receiver and those in the satellites due to poor synchronisation. Using the data from the fourth satellite, the receiver can find this timing offset and hence can eliminate it \cite{logsdon2012navstar,el2002introduction,januszewski2018gnss,ur2010characterisation}.\\
    
    \begin{figure}
        \centering
        \includegraphics[width=0.5\textwidth]{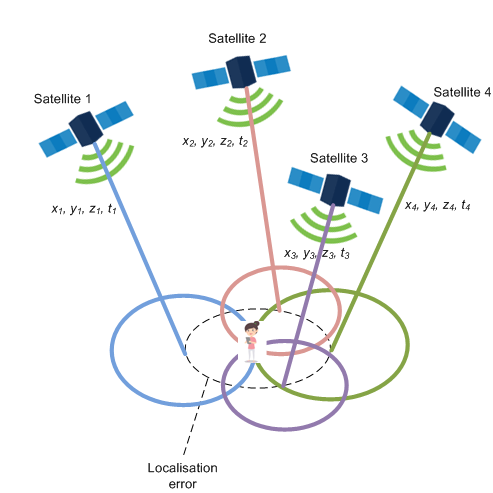}
        \caption{GPS Position calculation using triangulation}
        \label{fig_gps_1}
    \end{figure}
    
    GPS localisation is a fundamental tool in identifying the location of a smartphone user. By maintaining the database of geolocation information, people whose devices were in the same area in a certain time duration can easily be found. In the context of COVID-19, a user’s mobility history could be maintained for the last 14 days and securely stored on a restricted cloud server. Mobile phones whose owners have tested positive would be flagged on the app, and big data analytics would be used to determine which other phones have been in the proximity of that positive case within the historic window. Targeted messages could then be sent to phones that came in close contact, advising owners on whether to seek self-isolation or medical help. There would be minimal need for personal information as only the GPS location of the phone is required to identify the risk of exposure. Location and GPS data would also help officials to build maps of “transmission zones” that could paint a picture of how and where the disease is spreading Figure \ref{fig_gps_2}).\\
    
    \begin{figure}
        \centering
        \includegraphics[width=0.5\textwidth]{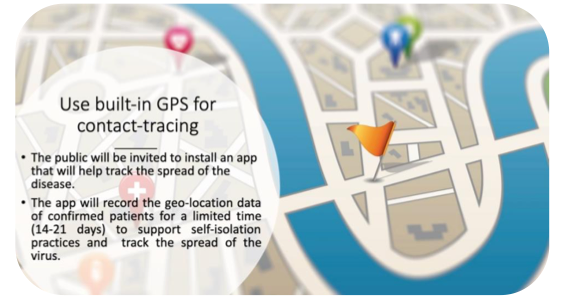}
        \caption{Active GPS tracking}
        \label{fig_gps_2}
    \end{figure}
    
    GPS tracking however is a significant drain on mobile phone batteries and is not accurate enough as GPS-enabled smartphones are typically accurate to within a 4.9 meter caused by signal blockage due to buildings, bridges, trees, etc., indoor or underground use and multipath reflection. Privacy is also another issue that restricts the wide usage of GPS for public location tracking. Another concern is spoofing attacks where a spoofer creates a false GPS signal with an incorrect time and location to a particular receiver \cite{dar2020}.\\
    
    \item \textbf{Bluetooth}: Contact tracing apps leveraging Low-power Bluetooth Communication (LBC) passively collect information about surrounding Bluetooth IDs by doing regular scans \cite{altuwaiyan2018epic}. The user grants the app access to the phone’s Bluetooth, which it uses to search for nearby Bluetooth devices (within 5-10 metre range). The phone then stores the list of Bluetooth devices it has encountered. Traditionally, a centralized approach is adopted where scans of individuals are uploaded to a central server database administered by health officials. Each scan includes the information of control flags, adjacent node ID, contact start time, contact end time and distance (discretized to ‘Close’, ‘Medium’, and ‘Far’ based on received signal strength indicator (RSSI) value) from the near contact. Pairwise matching scores between user data and the database are regularly calculated to identify contacts, whom a given user has been close to in the past 14 days. \\
    
    If a user turns COVID-19-positive, the list of Bluetooth devices encountered can be fetched, and the owners advised on whether to get tested or go into self-isolation. Additionally, Bluetooth beacons could be placed at specific locations such as grocery stores or in train coaches to determine which phones have visited those locations (at a specific date and time) \cite{cryptoeprint:2020:399}. This system can also alert venue managers to close or carry out a general sanitisation of the location/venue if a severe case or cluster of cases are identified \cite{avitabile2020towards}. Bluetooth addresses some disadvantages of GPS and expands the usability of the app with further qualities, as shown in Figure \ref{fig_bluetooth_1}.\\
    
    \begin{figure}
        \centering
        \includegraphics[width=0.5\textwidth]{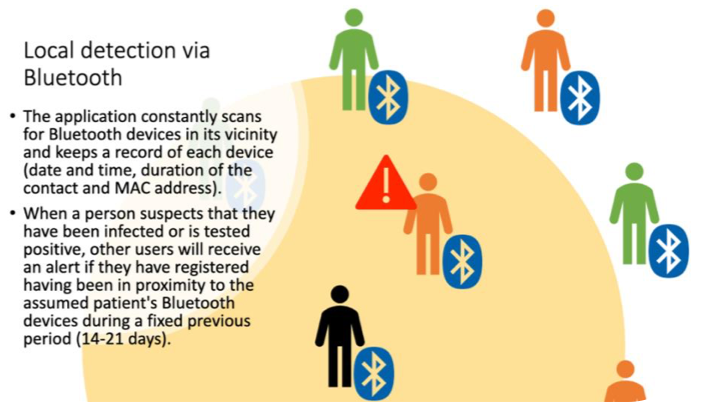}
        \caption{Local detection via Bluetooth}
        \label{fig_bluetooth_1}
    \end{figure}
    
    \begin{figure*}[t]
        \centering
        \includegraphics[width=0.9\textwidth]{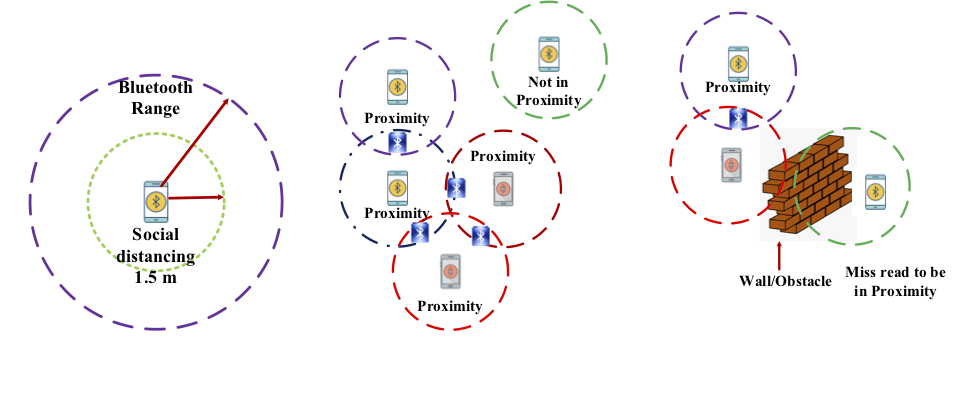}
        \caption{Proximity and not-in-proximity scenarios in Bluetooth contact tracing}
        \label{fig_bluetooth_2}
    \end{figure*}
    
    Systems using Bluetooth communication for automatic contact tracing has been first proposed by Altuwaiyan et al. in 2018 \cite{altuwaiyan2018epic}. However there are challenges to be addressed: Firstly apps like TraceTogether \cite{Tracetogether}, that works on similar idea of exploiting LBC for contact tracing are prone to information leakage due to centralized architecture; Secondly, the range of Bluetooth is more than 1.5 meters and can penetrate through walls, hence people in different rooms and behind other obstacles may also be regarded as being in contact generating false positives as shown in Figure \ref{fig_bluetooth_2}. Bluetooth addresses some disadvantages of GPS and expands the usability of the app with further qualities, as shown in Figure \ref{fig_bluetooth_2}. The effectiveness of Bluetooth-based approach however depends on the massive adaptability of the App \cite{Reuters_Bluetooth} and slow or low rate of adoption reduces its usefulness \cite{dar2020}.\\
    
    The privacy concerns can be addressed by incorporating cryptographic techniques to generate random keys and use them instead of phone ids' which are not just anonymous, but pseudonymous, constantly changing their ‘ID,’ and that cannot be tracked back to an individual \cite{NetworkWorld}. These contact-tracing keys will sit on your device, rather than in a centralised server. Moreover peer-to-peer decentralized contact tracing mechanisms \cite{brack2020decentralized} can be incorporated using distributed hash tables that makes use of blind signatures to ensure messages about infections are authentic and remain unchanged. Although Bluetooth can, with its wide range, detect another phone in its vicinity but cannot pinpoint which direction the contact is coming from. Ultra-wide band chips \cite{Wired} in the latest smart phones can help since it is possible to determine how close a person is by sending billions of pulses across a wide range of frequencies. However, the limitation is that not many phones have UWB (ultra-wideband) chips and they can only communicate with each other. \\
    
    \begin{figure*}[t]
        \centering
        \includegraphics[width=0.9\textwidth]{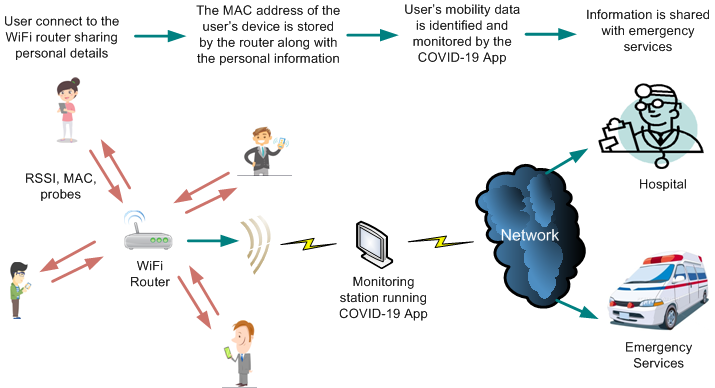}
        \caption{WiFi contact tracing approach}
        \label{fig_wifi}
    \end{figure*}
    
    \item \textbf{Wi-Fi Router Tracing}: Wi-Fi currently carries more than 60\% of the world’s Internet traffic \cite{Time}. Smartphone based localization methods employing wireless signals such as WiFi enjoy more popularity due to the use of off-the-shelf internal sensors, and relatively low cost \cite{huang2020wifi}. Ubiquity of Wi-Fi access through massive deployment of WiFi routers can be exploited to gain the knowledge of a user’s mobility data. These mobility traces are unique and identify the users accurately giving information about home and work locations, visited places, and personality traits \cite{de2013unique,de2013predicting}. High-resolution mobility patterns of entire social systems can perform an important role to ensure social distancing and combat the spread of epidemics including COVID-19 on multiple scales \cite{eubank2004marathe,sun2014efficient,liang2013unraveling,sapiezynski2015tracking}.\\
    
    When a smartphone user is within the range of a publicly available Wi-Fi and wishes to use it, it transmits a probe request to the router containing its globally unique Media Access Control (MAC) address. The router assigns a unique IP address to the user and maintain an entry into the Dynamic Host Client Protocol (DHCP) table. Most of the providers also ask for additional information such as name, email, location tracking, etc. to use it for business adverts. Using RSSI (Received Signal Strength Indication), MAC address, personal information and timestamps for the probe request of a user, the router can easily generate a WiFi signal map detecting the location of a user and duration of his presence based on the services used (Figure \ref{fig_wifi}). Cross referencing and basic analysis of logs from various routers will enable tracking in the locations a user visited, how long he spent in a specific area and how fast he moved from place to other. The average coverage range of the Wi-Fi is 80 meters outdoors and 50 meters indoors, limiting positioning to 4-15 meters. A combination of Wi-Fi with GPS is being used by Google, Apple, Microsoft, Skyhook, etc. to improve positioning \cite{de2013predicting}. \\
    
    \begin{figure*}
        \centering
        \includegraphics[width=0.9\textwidth]{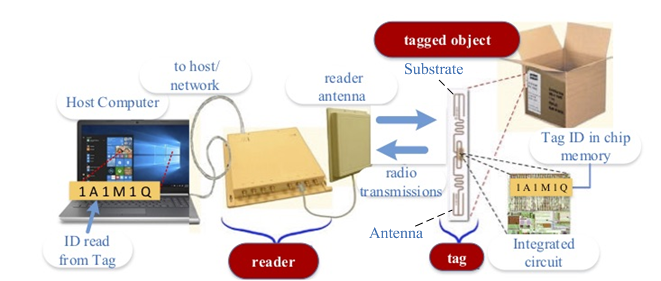}
        \caption{Key components in a typical RFID system}
        \label{fig_rfid_1}
    \end{figure*}
    
    The mobility data of an app user who self-diagnoses would identify the locations they have visited and people who have been in the vicinity (using the same router for Wi-Fi access). An alert would then be generated to advise these people to self-isolate.\\
    
    \item \textbf{Radio Frequency Identification (RFID)}: RFID is one of the major identification technologies used today covering almost every aspect of our daily life. Applications of this electromagnetic waves-based identification method include access to buildings and transportation, animal tracking, patient monitoring, personal identification, facilitating the inventory and shipping of goods, assembly lines and supply chains, tagging food and retail items, localization, and even providing assistance for visually impaired persons \cite{dobkin2012rf,finkenzeller2010rfid,chawla2007overview}.\\
    
    RFID technology has accomplished a major development in the last decade, mainly due to the reduction in the cost of RFID chips and huge developments in microelectronics and RF domains. Figure \ref{fig_rfid_1} illustrates a typical RFID System that comprises of a set of remote transponders known as RFID tags, and an RFID reader. RFID tags include an antenna and an application specific integrated circuit (ASIC) also known as a chip, containing the data about the tagged object. The RFID reader generates a query signal towards the RFID tags and the tag replies back with data. The Readers are usually connected with some embedded systems, host computers having application software to collect and share data. \\
    
    The passive Ultra High Frequency (UHF) tags typically consist of three elements; (1) transponder (packed in ASIC or simply an RFID chip); (2) antenna; and (3) dielectric substrate. Passive tags are usually very simple devices (Figure 6) and therefore, much cheaper (typically costing around \$0.10) than other types of radio devices. Passive tags do not require maintenance and have a long life, which is limited by the degradation of the label material rather than the use of batteries. Passive tags are expected to be readable for 10 to 20 years in many environments \cite{landt2005history,harrop2005rfid,das2010rfid,das2016rfid,das2018rfid,he2015rfid,al2015internet,griggs2018localizing,zanella2014internet}. \\
    
    \begin{figure*}
        \centering
        \includegraphics[width=0.9\textwidth]{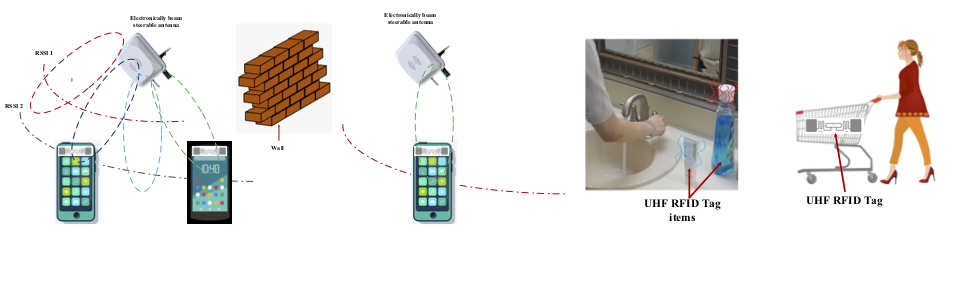}
        \caption{Reading the proximity of RFID tags using UHF RFID technology with beam steerable antennas}
        \label{fig_rfid_2}
    \end{figure*}
    
    UHF RFID uses passive tags attached to the smartphones and objects. It enables tracking of location of COVID-19 patient as well as items touched/used by him through Real Time Location Systems. The RFID reader antenna with beam steering capability will be used to read tag angle and RSSI, the tag proximity with other tags will be estimated by applying signal processing and machine learning techniques. RFID reader antenna will also not interfere with other RFID readers behind the walls/obstacle. A hybrid of Bluetooth and RFID can therefore be used to mitigate the drawbacks of Bluetooth and improve the accuracy (Figure \ref{fig_rfid_2}). The hybrid technique also has significant privacy advantages over GPS-based location-tracking. \\
    
    \item \textbf{UWB 5G Positioning}: 5G networks use large antenna arrays and ultra-wide bandwidths (UWB). They enable a decimetre level accuracy in location systems. Unlike other positioning technologies such as GPS, Bluetooth or WiFi, UWB technology uses RF signal’s Time Difference of Arrival (TDOA) or Time of Flight to estimate the distance between target and reference base station, that provides more accuracy with much more precise range measurement as shown in Figure \ref{fig_5G}. However, these systems are not yet fully operational. 
    
    \begin{figure}
        \centering
        \includegraphics[width=0.5\textwidth]{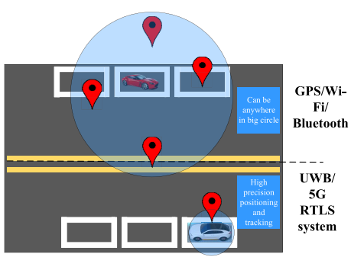}
        \caption{Difference between accuracy of 5G UWB and other positioning systems}
        \label{fig_5G}
    \end{figure}

\end{itemize}

\begin{figure}[ht]
        \centering
        \includegraphics[width=0.5\textwidth]{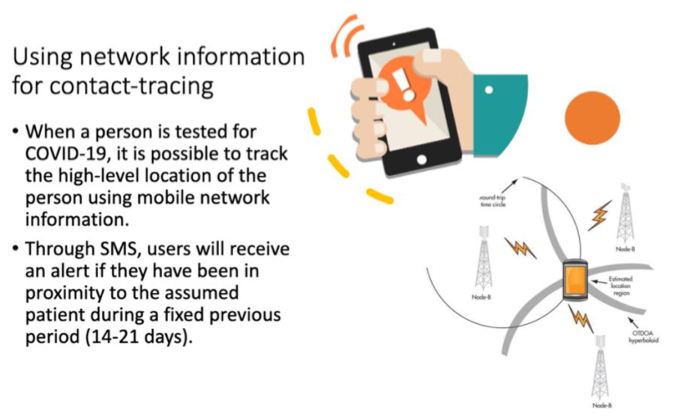}
        \caption{Contact tracing using network information}
        \label{fig_network}
    \end{figure}

\subsection{Mobile Network Tracing}
Mobile network operators already hold information on subscriber location and mobility history, albeit the resolution of the subscriber localisation is only down to the base station level. However, with the growing trend of small cells covering urban centres and the implementation of Artificial Intelligence techniques, this mode of tracing could be efficient and useful. This solution involves using mobile network information and radio control signals to get the user’s location and mobility history, as shown in Figure \ref{fig_network}. This information can be used for high-level epidemiological studies to determine the spread of the disease especially after crowded events such as religious gatherings, parties, concerts or sporting events (likely to open gradually after the lock-down is progressively relaxed). Text messages will be sent to subscribers asking them to opt out if they do not wish to participate.

\begin{figure*}
        \centering
        \includegraphics[width=0.7\textwidth]{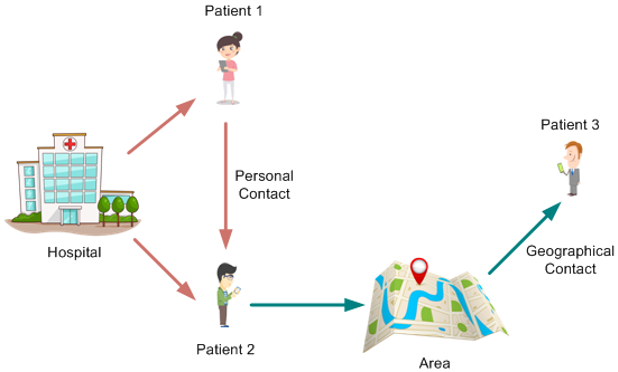}
        \caption{Example of contact network creation using two-mode network approach}
        \label{fig_network_2}
    \end{figure*}

\subsection{Crowd Sourcing of Social Media}
Social media analytic can be expanded by fusing together additional data sources such as License Plate Recognition (LPR), smart city CCTV, ATM transactions and credit card purchases, to help recreate the possible corona virus exposure path \cite{karisani2020mining,chen2011social,CovidGraph,garcia2014using}. Leveraging graph database and graph inference algorithms, we can model complex interactions of individuals/group of individuals by linking and correlating information from heterogeneous digital data sources (online activity and check-ins, ATM transactions and LPR to detect visited locations, geolocation information inferred from mobile phone data or WIFI tracing) \cite{xu2020twitter}. In these types of specialized databases, people, places, and things are treated as “nodes” and the connections between them are called “edges” that makes up the COVID-19 contact network. These networks consisting of nodes and edges make it possible to illuminate collected knowledge clearly, to uncover connections and to recognize patterns, and graph analytic can be used to detect contacts of infected people (clusters). A contact network is not necessarily a social network, since contacts might be family, friends, acquaintances, or strangers. We can pick up a disease from a sick family member at home or from a stranger via an inopportune sneeze in a crowded coffee shop.

Human networks evolve into what social scientists call “small-world networks” - we tend to cluster together via the social dynamic of hemophilia (i.e. birds of a feather flock together). A small-world network is made up of connected clusters where there are more connections within the cluster than between the clusters. Shortest path algorithms can also be used to trace infection paths across multiple contact points within clusters, thus revealing disease pathways. Identifying ‘super-spreaders’ or super-spreading events among the infected patients can be conveniently studied and visualized from a graph network perspective. Two-mode network approach \cite{chen2011social} can be adopted to create COVID-19 contact network such as shown in Figure \ref{fig_network_2}, consisting of different layers, personal and geographical which emphasizes the relationships between other individuals and their visits to high-risk locations. This would enable proximity tracking of all the other individual/group of individuals who are most likely be in contact with the carrier or are present in the same time and space.

\begin{figure*}
        \centering
        \includegraphics[width=0.9\textwidth]{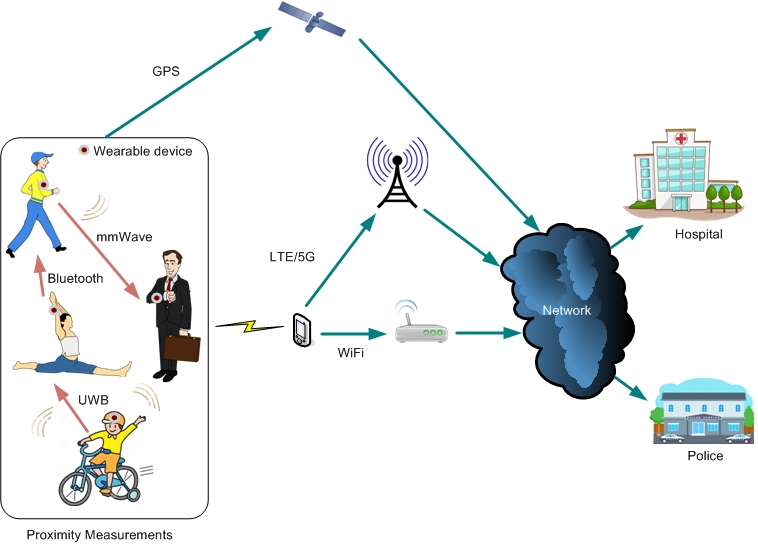}
        \caption{Use of wearables for contact tracing}
        \label{fig_wearables}
    \end{figure*}

\subsection{Wearable Devices}
Wearable devices enable mobile computing and wireless networking and collect data, track activities, and provide customized experiences to the user's needs and desires. Wearable technology has successfully moved past the adoption stage and now stands at the brink of massive diversification with an explosion in popularity and applicability. Wearable devices have found applications in almost every aspect of our daily life including consumer electronics, healthcare, sports and fitness, enterprise and industry, media and entertainment \cite{abbasi2016advances}. Smart glasses, smart watches, wristbands, fitness and health trackers, smart clothing and smart jewellery are the most popular wearable devices. The global market of the wearable technology was estimated at \$24.6 billion in 2019 and is expected to hit the \$38.41 billion mark by 2025 \cite{MarketWatch}. It would cause the wearable devices market to grow from 216 million devices in 2019 to 614.3 million units in 2025 \cite{MordorIntelligence}. 

Wearables make use of a number of technologies including cellular, Near Field Communication (NFC), Bluetooth, Wi-Fi, GPS, ultra-wideband (UWB), Long Term Evolution (LTE) and 5G for information gathering, communication and localisation. A typical wearable device can perform the contact tracing feature for COVID-19 by tracking the location of the user through GPS tracker as well as proximity sensors utilising Bluetooth, UWB radio and LTE/5G connectivity. When a user becomes COVID-19 positive, they updates their status that prompts alerts to others who have been in contact with them based on proximity and location data history. The wearable devices can also be used effectively to enforce the self-isolation/quarantine through monitoring of geolocation of the user and raising alarm when such a person moves out of their house. Figure \ref{fig_wearables} shows a possible working scenario of the wearable devices for contact tracing. Use of multiple technologies makes wearables quite versatile in terms of performance, range and accuracy.  

Health data like heart rate, skin temperature, cough and blood oxygen saturation collected from wearable devices can also be effectively used to detect the onset and progression of illness caused by COVID-19 \cite{viboud2020fitbit}. By leveraging wearable technology either off-the shelf solutions like FitBit or customized solutions (electronic bracelets equipped with vital signal monitoring, geolocation sensor, proximity sensors powered by Bluetooth, GPS), contact tracing can be enhanced by connecting it to a mobile app such as StayHomeSafe \cite{StayHomeSafe}. Scripps Health, Stanford Medicine and Fitbit are collaborating in a study to assess the ability of wearable devices to track, trace and isolate COVID-19 patients \cite{HealthcareITNews}.

\begin{figure}
        \centering
        \includegraphics[width=0.5\textwidth]{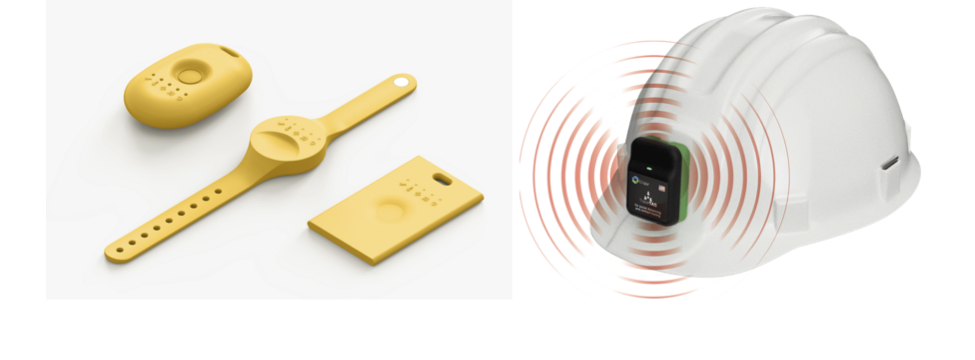}
        \caption{Examples of wearables for contact tracing \cite{TechCrunch,PowerMag}}
        \label{fig_wearables_2}
    \end{figure}

Estimote has created a range of wearable devices called the “Proof of Health” wearables for COVID-19 contact tracing at the level of the local workplace \cite{Estimote,TechCrunch}. The device that can be worn using a lanyard or like a wristwatch (Figure \ref{fig_wearables_2}), has passive GPS location tracking, Bluetooth and UWB proximity sensors, a rechargeable battery and built-in LTE. A change in the user’s health status, such as COVID-19 suspected or confirmed, flags the alarm and alert the people who have been in contact with them. This however requires use of Estimote’s wearable devices by the whole work force in that facility.

The Proximity Trace device developed by Triax Technologies, Inc., is affixed to a safety hat or worn on the body with a lanyard alerts the user when they are too close to another user. In case of a COVID-19-positive user, the company can perform contact tracing and identify possible exposures using historical data logged in passively by the user’s device \cite{PowerMag}. Similar solutions are also offered by Blackline Safety Corp \cite{BusinessWire}. A contact tracing App, Workforce that is compatible with the wearable wrist bands for contact tracing and tracking is developed by Ascent Solutions \cite{MyAscents}.

Wearable Bluetooth/GPS enabled wristbands have been tested for contact tracing, enforcing the social distancing and observing the lockdown during COVID-19 pandemic in countries including Bulgaria, South Korea, Hong Kong, Belgium, Lichtenstein and India \cite{BBC_WristBands}. 

These wearable devices must be synced to their home location through smartphone’s GPS and any active patients can be tracked and made to remain in that location until cleared. An alert is sent to the monitoring station if the wearer moves further than 15 metres away from their phone. The location and proximity data of such individuals can be mapped to a centralized web-based dashboard as a warning mechanism or can be coupled with centralized tracing applications like TraceTogether \cite{Tracetogether}.

\renewcommand\theadfont{\bfseries}
\begin{table*}
    \caption{Comparison of COVID-19 contact tracing enabling technologies}
    \centering
     \setlength{\leftmargini}{0.4cm}
    \begin{tabular}{| m{2.5cm} | m{5cm} | m{5cm} | m{1.5cm} |}
        \hline
        \thead{Technology} & \thead{Pros} & \thead{Cons} & \thead{Users} \\
        \hline
        GPS & 
        \begin{itemize} 
            \item Availability of real-time location information. 
            \item Locating users in real-time who contracted the virus.
            \item Identifying the demand and need for healthcare in an area.
            \item Identify virus hot-spots with Geo-data.
            \item Local information and awareness for patients, carers.
            \item Enabling care professionals to continue the upmost service in care.
        \end{itemize} & 
        \begin{itemize} 
            \item High storage and computational requirements. 
            \item Possible issues with indoor localisation. 
            \item Social fears of being tracked and the lack of trust in the use of personal/health. 
        \end{itemize} & 
        Everyone \\
        \hline
        Bluetooth & 
        \begin{itemize} 
            \item Wide availability. 
            \item Low power requirements.
            \item List of all devices that have “made contact” is readily available.
            \item Reduced requirement for storage and computational resources. 
        \end{itemize} & 
        \begin{itemize} 
            \item Inaccuracy of proximity approximation.  
            \item 5-10 meters scanning range causing false positives.
            \item Real-time location information is not available.
            \item Requires a higher level of programming to make sure the Bluetooth connectivity is enabled and responsive to the requirements of the application. 
        \end{itemize} & 
        Everyone \\
        \hline
        Bluetooth plus UHF RFID & 
        \begin{itemize} 
            \item Able to track the contacts with accuracy and double check. 
            \item Can track the belongings and items in use of the patient.
        \end{itemize} & 
        \begin{itemize} 
            \item Tagging items and deploying RFID readers with phase RSSI and phase reading will incur cost. 
             
        \end{itemize} & 
        Everyone \\
        \hline
        WiFi router tracing &
        \begin{itemize} 
            \item Widely available worldwide as handling 60\% of the internet traffic. 
            \item Wide range of different types of existing WiFi routers can be readily used with no extra hardware.
        \end{itemize} & 
        \begin{itemize} 
            \item Relatively low accuracy. Hybrid techniques such as using the built-in accelerometer and gyroscope with WiFi can improve the accuracy. 
            
        \end{itemize} & 
        Everyone \\
        \hline
        Mobile network tracing & 
        \begin{itemize} 
            \item No app installation is required.
            \item Transparent to the user.
            \item Larger public access, who, if desired, could opt-out of the programme.
 
        \end{itemize} & 
        \begin{itemize} 
            \item List of devices that have made contact is not available.
            \item Only high-level localisation information is available.
            \item Participation of network operators is required to increase coverage.

        \end{itemize} & 
        Mobile Operators \\
        \hline
        UWB 5G  & 
        \begin{itemize} 
            \item High accuracy.
        \end{itemize} & 
        \begin{itemize} 
            \item Not yet fully operational. 
        \end{itemize} & 
        Everyone \\
        \hline
        Crowd sourcing of social media and tracking financial transactions  & 
        \begin{itemize} 
            \item A pre-outbreak pattern can be identified indicating the areas where the virus could strike next.
            \item Could generating near-real-time information for public health officials that could help tracking its spread.

        \end{itemize} & 
        \begin{itemize} 
            \item Privacy is a major concern as the accuracy of such models depends on the location information and other data sources including financial transaction information. 
        \end{itemize} & 
        Government entities \\
        \hline
        Wearable devices using Bleutooth/GPS/WiFi & 
        \begin{itemize} 
            \item Suitable for high traffic and dense areas including indoors, malls, homes.
            \item Enables tracking as well as geofencing the infected patients.
            \item Allows remote tracking of quarantined patients.
            \item Increased coverage and reliability.
            \item No need of a smartphone, hence cost effective.

        \end{itemize} & 
        \begin{itemize} 
            \item Requires a customized wearable device. 
            \item Requires the user to have the app running at all times.
            \item Can generate false positives as range is higher than 1.5 meter.
 
        \end{itemize} & 
        Everyone \\
        \hline
    \end{tabular}
    \label{tab_technology_comp}
\end{table*}

\section{Attacks}
There are a range of attacks on contact tracing method. As contact tracing is heavily reliant on the use of data science and machine learning, it inherits the vulnerabilities of those areas. Let us consider a systematic monitoring of people's proximity, which results in a classification mechanism depending on that proximity if there is high probability for an infection. A potential instance of the recorded interactions is a sample that is classified using \textcolor{black}{utility function} $f$ either as probably infected or not infected. Let us assume that we have an input space $\mathcal{X}=\{x_i\}$ and of course an output space $\mathcal{Y}=\{y_i\}$, where $x_i$ is an instance of the interactions and $y_i$ is the output of that instance determined by $f$, i.e. $f(x_i)=y_i$. We make the assumption that our initial system has been trained using $N$ samples from the real world where we know the interactions of people and how many of them were eventually infected. Those samples form the training set $\mathcal{S}$ and it has derived the \textit{system perception}, denoted by $\hat{y}$. After the end of \textcolor{black}{the training phase}, our designed system receives new samples from the real world environment and classifies them. We are able to define this as the \textit{run-time phase} of our system. For every new event $\hat{x}_i$, $f$ gives a new output $f(\hat{x}_i)=\hat{y}_i$.
We have the following cases:

\begin{itemize}
    \item If $\hat{x}_i$ are probably infected and our system does not recognize them as such, they are called false negatives that cause a loss $ l$  to our system.
    \item If $\hat{x}_i$ are probably infected and our system recognizes them as such, they are called true positives. They might also be not infected. In either case there is no loss to our system.
    \item If $ \hat{x}_i$ are not infected and our system recognizes them as probably infected, they are called false positives and cause a loss $\lambda$ to our system.
\end{itemize}

The attacker of the specific system wants to avoid being self isolated because of their contact with infected people. The aim of the attacker is to maximize the impact the attack has to the system by maximizing $|f(\hat{x}_i) - y_i|$. Consequently, a challenge of the system that defends its functionality is to find a utility function that minimizes the losses, measured as the distance of $f(\hat{x}_i)$ to the real output $y_i$.

\subsection{Attacks on Data Privacy}
\label{dataprivacyattacks}

\subsubsection{Membership Inference Attacks}
An adversary is able to perform a membership inference attack \cite{shokri2017membership} by querying the model and exploiting the returned confidence values to distinguish if data was part of the training. In a contact tracing application, the returned output to the users would include if the person has been in contact with a COVID-19 patient, alongside with a confidence score of this classification. An adversary is able to exploit this information, with knowledge extracted from public data sets, to identify if people's data was part of the ML training.

\subsubsection{Model Inversion Attacks}
\label{modelinvattacks}

In Model Inversion (MI) attacks \cite{fredrikson2014privacy,fredrikson2015model,zhang2019secret}, the adversary's goal is to reproduce the sensitive training data. The threat model of MI attacks involves access to the training model, as well as the confidence scores that are returned as the output of it. In the contact tracing App, the ML model should be hidden from the users, since there is no need for them to access it.

\subsubsection{Data Poisoning Attacks}
\label{datapoisoningattacks}

The adversary can poison the training data set. To accomplish their goal, they \textit{derive} and \textit{inject} a point to decrease the classification accuracy \cite{steinhardt2017certified,laishram2016curie,biggio2012poisoning,jagielski2018manipulating,munoz2017towards,yang2017generative}. This attack has the ability to completely distort the classification function during its training, thus allowing the attacker to divert the classification of the system according to their taste. In a real-world scenario, the attacker needs to have privileged access to a contact tracing application, being a member of either the development or the administration team. This scenario is rather unrealistic as a single person would not be responsible for handling the training data set without any supervision.

\subsection{Attacks on Model Privacy}

\subsubsection{Evasion Attacks} 
The attacker can undertake an \textit{evasion} attack against classification during the testing phase, thus producing a wrong system perception. In this case, the goal of the adversary is to achieve misclassification of their data, for example, remaining unnoticed. In a real-world scenario with a contact tracing app, the attacker would want to confuse the system about the interactions they had with other people. Moreover, the attacker can compromise the targeted system by being spotted out as potentially infected. This can be easily achieved by using many different phones in too many different locations. There was an incident where a person caused an artificial traffic jam on Google maps using a wagon full of Phones \cite{howtofake_googlemaps}. This incident changed the perception of the system about road traffic, forcing Google maps to assume that there was a traffic jam when it was not busy.

\subsubsection{Model Extraction Attacks}
The adversary in Model Extraction attacks \cite{tramer2016stealing} is trying to reconstruct a ML model that is similar to the original, by identifying the decision boundaries of it. The attackers aim to have complete access to a ML model that behaves similarly with the original, in order to perform an attack on Data Privacy, as seen in Section~\ref{dataprivacyattacks}. 

\subsubsection{Model Poisoning Attacks}
{Model Poisoning attacks are quite similar to the Data Poisoning attacks (discussed in Section~\ref{datapoisoningattacks}) as an adversary injects ``hidden'' poisons to the training model in order to behave maliciously only on their trigger \cite{gu2017badnets,liu2017trojaning}.} However, Model Poisoning attacks, opposed to Data Poisoning attacks, do not require access to the ML training procedure, and elaborate scenarios where the ML model is sent to the users, such as Federated Learning \cite{mcmahan2016communication,mcmahan2017federated,konevcny2016federated,bonawitz2017practical}. In the contact tracing scenario, the ML model can be sent to the users for training, and then to protect the privacy of the users, a secure aggregation technique collects and aggregates all the trained models, before returning it to the ML coordinators. That enables the Model Poisoning threat possibility, since a malicious user can poison the received model before sending it back, and since the secure aggregation is in place, the ML coordinators can not identify that the final trained model is being poisoned.




\section{Critical Analysis}
The discussion on user and contact tracing opens up a whole lot of questions, and the most fundamental of these is that we do not actually have any real infrastructure to implement privacy-preserving methods. {It is likely that a COVID-19 App would be a pin-point App where the data gathered for location and contact tracking could be easily abused. Furthermore, in the absence of a universally acceptable standard, it would have limited scope outside a country’s borders.} Our major problem is that we have built data infrastructures that mirror those from the 1980s where we care little about the core rights of the data we gather. Once captured, the owner becomes the entity who captured the data, and without the trustworthiness of the transactions involved, we leave it open to abuse for malicious activities.

\subsection{Who is Trent?}
\begin{figure*}
\centering\includegraphics[width=1.0\textwidth]{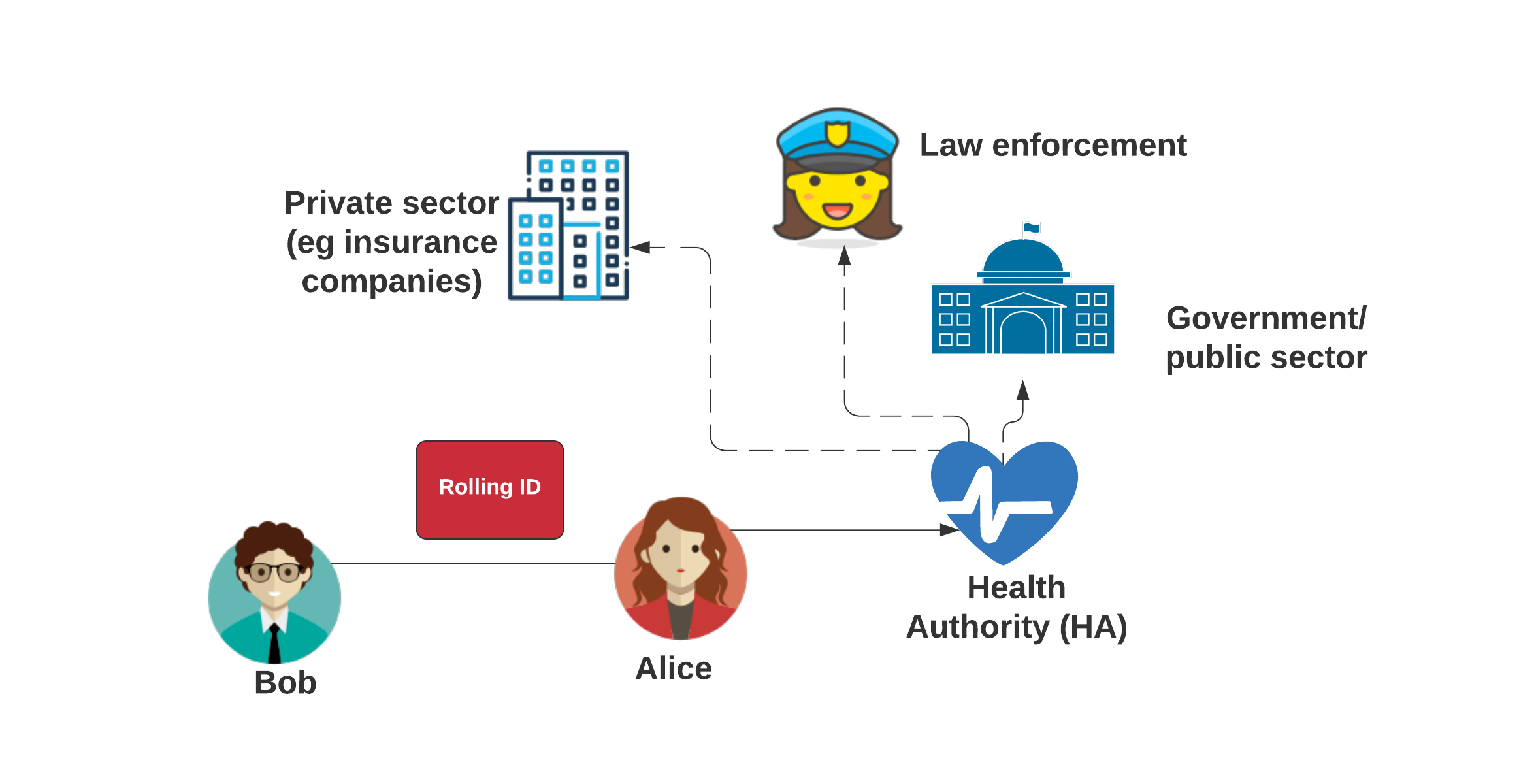}
\caption{Who is Trent?}\label{fig_trent}
\end{figure*}

We often trust our health authorities a great deal more than we trust our governments. If the data goes to clinical staff for analysis, we perhaps find that more acceptable than someone in law enforcement. Thus, the fundamental question in the whole system is how we can make sure that Trent is someone trusted. This could be a trusted entity who handles the data on behalf of the citizen (and preserves their privacy on their behalf) or a health authority that Alice trusts? It is unlikely that we should trust anyone {other than health authorities, and also we need to make sure that the data gathered is only kept for the required amount of time. Any tracing of contacts should not be kept for longer than it is required, it should be used only for the clinical purposes and only provided to trusted health professionals (Figure \ref{fig_trent}).} 

\subsection{Where is Carol?}
{There are no trusted mechanisms to integrate a formal test for COVID-19. we therefore bring in Carol the Tester.} Carol the Tester would be able to define a state of testing: Positive, Clear, and Suspected, and will have followed a scientific process to provide Alice’s COVID-19 status.

\subsection{What is Carol’s Attestation?}
{We don’t want the government to control testing, making availability of a trusted network of testers — from our own country and from other trusted places, eminent. Thus, we need a trusted way for a number of testers to sign the attestation that Alice has COVID-19, or when she has been identified to be negative. A new attestation will revoke a previous one. So, will this attestation be in the form of passporting system with digital verification allowing Alice to carry a digital passport of being free from COVID-19, and that can be passed to others?}

\subsection{Clear Role for Trent}
{Overall, Trent must be the health authority and not the government. Trent must be the one who marks the status of Alice as: Clear, Suspected, Not Known, and Positive. In this way, Trent is responsible for defining the status of Alice and the tracking will only happen when there is a positive status applied to her.}

\subsection{Trent Must be the Health Authority}
Border control, and which countries/testers do you trust? The basic flaw of having just one health authority involved is that there needs to be a trust between health authorities in different countries. In this way, we can trace over borders, when the travel {reopens. A trust network for Carol in each country needs to be defined along with the definition of attestation process so that it can be shown at borders. Countries thus will define the testing trust network, where Carol’s testing is acceptable cross-border.}

\subsection{There is No Real Integration of Identities}
How we are going to properly {identify} Bob, Trent or Alice? With having a private key signing for Alice and Trent, there is no real way of knowing that Alice is Alice and Trent is Trent. Thus, we need someone for each core identity to be certified for the things they are signing. This might include the health authority to self-certify.

\subsection{There is No Clear Mechanism for Alice’s Consent}
A fundamental flaw is the consent mechanisms that Alice must give to Trent in order to flip her current COVID-19 statement, especially in the changing it to a positive state. While Trent may have the rights to record Alice as being positive, he may need Alice’s consent as to whether she is okay with her COVID-19 state being broadcast to others.

Normally when an App is installed on a mobile device, the user should be alerted to the privacy implications of the use of their personal data and would then need to give their consent before proceeding. The consent would in this instance cover the transmission of personal data to health authorities or other official bodies. Under the EU General Data Protection Regulation (GDPR) it could be argued that vital interests of individuals would be an alternative legal basis for gathering and sharing this personal data \cite{EuropeanParliament2016}. This would apply to life-or-death situation and this could be difficult to argue solely on the basis that there is a possibility of death following exposure to an infected individual. The other legal basis for processing personal data would be public interest – and this might be more easily justified on the basis that protecting public health is in the public interest. Following GDPR principles will be particularly important as the UK is likely to continue to be subject to EU law during the pandemic. 

\subsection{Long Term Data Retention}
The UK's Data Protection Act 2018 allows for personal data to be kept for research purposes so long as reasonable steps are taken to protect individual identity \cite{UKParliament2018}.  Pseudonymisation may offer some protection, but is not a guarantee that privacy will be preserved. For instance, the data could be combined with published data sets such as electoral registers to identify individuals.

\subsection{Lack of Incentive for Uptake}
{To make the App a success, there should be strong incentives for uptake.}  It may not be sufficient to appeal to public-spirited attitudes when there is a strong disincentive of increased personal restrictions. If an individual receives an alert that they have been in close proximity to a suspected infected individual, they are advised to self-isolate. This will depend on the criteria that are applied to different levels of alerting, the accuracy of the proximity measures and the degree of self-reporting.  It is possible that incentives such as legal sanction and law enforcement are more likely to work in authoritarian states, whereas social pressure might work better in mono-cultural conformist societies. For a pluralistic and democratically accountable society offering privileges such as access to travel or greater freedom of movement might be better motivators. 

\subsection{Poor Data Quality}
Data quality will affect the contact tracing approach in several ways:
{Where it depends on self-reporting of COVID-19 symptoms, not everyone will do so}.  Will the authorities act on self-reporting in order to respond rapidly, or wait for the outcome of COVID-19 tests to confirm the self diagnosis{?} The follow-up tests are not completely accurate and have varying levels of type 1 and type 2 errors.  The potential emergence of a new, more virulent mutant strain of COVID-19 may affect the criteria used to assess risk (potential to transmit in less time or at an increased distance, for instance) \cite{Korber2020}.

\subsection{No Binding of Data}
There is no method which truly binds the capture of contact data to the entities involved in a trusted way. The only thing it does is to mask Alice until she is proven to have COVID-19, after which her data can be revealed to others without any restrictions.

\subsection{There’s No Signing Involved}
{A fundamental trait of the modern world is that we introduce proper digital signing. As a minimum, we should see key pairs being created for devices and entities where IDs and tracking are signed by private keys, and checked for the correctness of the signer. There are many trusted signing methods which can be used to anonymise the signer.}

\section{Conclusions}

We have identified a number of existing technologies including GPS, Bluetooth, Wi-Fi, RFID, wearable devices and social media fingerprints that can effectively provide the tracks of a COVID-19 patient. The potential contacts/proximity users can efficiently be identified and notified of the threat and hence, advice to self-isolate. Table \ref{tab_technology_comp} summarises their key benefits along with the disadvantages. 

The Bluetooth approach, being pursued at various stages by governments across Europe and Latin America, as well as in Australia and many Asia nations, requires a majority of people in a geographic area to adopt it for it to be effective. These apps are also considered to be interfering with vital signs monitoring applications such as diabetes monitoring \cite{Biggs_COVIDSafe}. Some countries, including South Korea and Israel, are using high-tech methods of contact tracing that involve tracking peoples’ location via phone networks. But such centralized, surveillance-based approaches are viewed as invasive and unacceptable in many countries for privacy reasons. 

The Bluetooth-based Apps are also more privacy-friendly than tracking techniques that use GPS or cellphone data. They use Bluetooth to broadcast and receive an encrypted, pseudonymous signal from nearby phones and create a log of interactions that remain on the phone, so users’ names and numbers are not disclosed. Social Media approach also has good potential but is marred by authentication restrictions. Wearables appear to be a dynamic and effective solution as have the capability to make use of multiple technologies with improved efficiency and higher accuracy.

The main issues that surround these enablers of a potential contact tracing application include privacy concerns, security loopholes, lack of testing and part use of the smartphones. The privacy concerns need to be eradicated through GDPR compliance, transparent development of the app and data usage and reassurance about the temporary nature of the surveillance.

\section*{Acknowledgements} {
}
The authors also acknowledge inputs and discussions from the following colleagues: Dr Ahmed Zoha, Dr Yusuf Sambo, Dr Shuja Ansari and Dr Kia Dashtipour. We acknowledge the funding for the Scotland 5G Centre which has partially supported the work behind this document.

\bibliographystyle{IEEEtran}
\bibliography{references}

\end{document}